\documentclass[12pt]{article}

  \usepackage{graphicx}
  \usepackage{epsfig}
  \usepackage{amssymb}
  \usepackage{subfigure}
  \usepackage{amsmath} 

\evensidemargin - .5truecm
\allowdisplaybreaks[1]

\def \lsim
{\raisebox{-3pt}{$\>\stackrel{<}{\scriptstyle\sim}\>$}}
\def \gsim
{\raisebox{-3pt}{$\>\stackrel{>}{\scriptstyle\sim}\>$}}

\begin{document}

\newcommand{\CC}{{\mathbb C}}
\newcommand{\RR}{{\mathbb R}}
\newcommand{\ZZ}{{\mathbb Z}}
\newcommand{\QQ}{{\mathbb Q}}
\newcommand{\NN}{{\mathbb N}}
\newcommand{\beq}{\begin{equation}}
\newcommand{\eeq}{\end{equation}}
\newcommand{\beal}{\begin{align}}
\newcommand{\eeal}{\end{align}}
\newcommand{\nn}{\nonumber}
\newcommand{\bea}{\begin{eqnarray}}
\newcommand{\eea}{\end{eqnarray}}
\newcommand{\ba}{\begin{array}}
\newcommand{\ea}{\end{array}}
\newcommand{\bfig}{\begin{figure}}
\newcommand{\efig}{\end{figure}}
\newcommand{\bc}{\begin{center}}
\newcommand{\ec}{\end{center}}

\newenvironment{appendletterA}
{
  \typeout{ Starting Appendix \thesection }
  \setcounter{section}{0}
  \setcounter{equation}{0}
  \renewcommand{\theequation}{A\arabic{equation}}
 }{
  \typeout{Appendix done}
 }
\newenvironment{appendletterB}
 {
  \typeout{ Starting Appendix \thesection }
  \setcounter{equation}{0}
  \renewcommand{\theequation}{B\arabic{equation}}
 }{
  \typeout{Appendix done}
 }

%
%
%
%

\begin{titlepage}
\nopagebreak

\renewcommand{\thefootnote}{\fnsymbol{footnote}}
\vskip 2cm
\begin{center}
\boldmath

~\\ {\Large\bf Improved Factorization for Threshold  \\ [0.2cm]
Resummation in Heavy Quark to Heavy Quark Decays}

\unboldmath
\vskip 2cm
{\large  U.G.~Aglietti${}^{(a)}$ and G.~Ferrera${}^{(b)}$}
\vskip .5cm
{\it ${}^{(a)}$ Dipartimento di Fisica, Universit\`a di Roma ``La Sapienza'',\\ I-00185  Rome, Italy}
\vskip .2cm
       {\it ${}^{(b)}$ Dipartimento di Fisica, Universit\`a di Milano and\\
         INFN Sezione di Milano, I-20133 Milan, Italy}
\end{center}
\vskip 1.7cm

\begin{abstract}
\normalsize
We consider the resummation of soft-gluon effects
in heavy quark to heavy quark decays,
namely the processes 
$Q_1 \to Q_2 \, + \, \mathrm{(non\,QCD\, partons)}$,
where $Q_1$ and $Q_2$ are two different heavy quarks.
We construct a new factorization scheme
for threshold resummed spectra,
which allows us to consistently evaluate
the distribution of the final hadron invariant mass $m_X$
in all the kinematic regions, i.e.\ when $m_X$ is smaller,
of the same order, or larger than the mass of the final
quark $Q_2$. 
A dependence of the Improved Coefficient Function
on the threshold variable is introduced,
which can however be relegated to a small
interval of this variable by means
of the so-called Partition of Unity. 
We explicitly apply our improved scheme to the
$b \to X_s \, + \gamma$ decay at next-to-leading logarithmic accuracy.
\vskip .4cm

\end{abstract}
\vfill
\end{titlepage}    

\setcounter{footnote}{0}

\newpage


\section{Introduction}

Analytic studies of Quantum-Chromo-Dynamics $(QCD)$
in the physical case, i.e.\ in four space-time dimensions, 
are substantially restricted to perturbation theory.
The systematic application of perturbation theory
to $QCD$ gives rise to the well-known
perturbative Quantum-Chromo-Dynamics ($pQCD$),
a mature branch of theoretical physics,
deeply involved in the phenomenology of the Standard Model. 
While exact analytic solutions of $QCD$ 
correlation functions could be
intrinsically beyond human ability 
\cite{Aglietti:2018aqc},
perturbative calculations
of cross sections and decay rates often reveal 
a rich structure 
and describe a variety of physical effects. 

There are basically two different approaches in $pQCD$.
The first one involves an {\it exact} evaluation
of the $QCD$ matrix elements 
of the process under investigation,
up to a given order $n$
in the coupling $\alpha_S$ ($n$ is taken, of course, as large as possible):
\beq
\sigma \, \simeq \, \sum_{k=0}^n \alpha_S^k \, \sigma^{(k)},
\eeq
where $\sigma^{(k)}$ in the contribution 
to the physical cross section $\sigma=\sigma(\alpha_S)$ of order $k$.
That way, all physical effects which show up in the matrix elements 
up to the truncation order $n$ are trivially taken into account.
In this approach, one has to assume that the
higher order $(k>n)$ contributions
to the cross section can be safely neglected.
In practice, one has to assume that all the
terms in $\sigma^{(k)}$ are of order unity.


In the second approach, one concentrates instead,
from the very beginning,
on a specific physical effect.
Usually, such effect manifests itself in peculiar
perturbative terms, which depend on a kinematic parameter,
and which become large in some region
of the space of this parameter.
Actually, at a generic order $k$, such enhanced terms $s^{(k)}$, 
contained in $\sigma^{(k)}$,
can become so large as to cancel the smallness of $\alpha_S^k$, the 
$k$-th power of the $QCD$ coupling. 
We face the situation
\beq
\alpha_S \, \ll \, 1, \qquad \left| s^{(k)} \right| \, \gg \, 1,
\eeq
with:
\beq
\left| \alpha_S^k \,  s^{(k)} \right| \, \gsim \, 1,
\quad \mathrm{for } \,\, 
k \, = \, 1, \, 2, \, 3, \, \cdots.
\eeq
For concreteness sake, we have assumed that 
the smallest order at which the physical effect under 
consideration manifests
itself is one, i.e.\ 
\beq
k \, \ge \, k_{\min} \, = \, 1,  
\eeq
but higher values of $k_{\min}=2,3,\cdots$, are possible; 
Usually, $k_{\min}$ is a small integer.
In this physical situation, it is necessary to resum
the enhanced terms $s^{(k)}$ to all orders of perturbation
theory, i.e.\ for all $k$. 
Schematically, we can write:
\beq
\sigma \, \simeq \, \sigma^{(0)} \,+\, \sum_{k=1}^\infty \alpha_S^k \, s^{(k)}. 
\eeq
Therefore, in the second approach, we realize an 
{\it approximate resummation} of the perturbative series
of the given cross section to all orders.
Note that, at each perturbative order $k$, 
we do not compute the exact
cross section contribution $\sigma^{(k)}$, 
but only the {\it leading} term 
$s^{(k)}$ contained in it,
\beq
s^{(k)} \, \simeq \, \sigma^{(k)},
\eeq
as far as the physical effect under consideration
is taken into account. 

From the above considerations, it should be clear that
fixed-order calculations and resummed ones 
rely on quite different philosophies.
In order to obtain an {\it optimal} perturbative description
of the process, one should combine
in some way the two approaches.
That involves the so-called matching operation 
--- or simply matching --- in which one
requires consistency between the two approaches. 
Roughly speaking, one would like to have an improved perturbative
formula for the cross section which, at low orders, contains
the exact matrix elements while, at higher
orders, contains the approximate matrix elements of the resummation.


In this work we consider the general process
\beq
\label{general_process}
Q_1 \, \to \, Q_2 \, + \, \mathrm{(non \, QCD \, partons)},
\eeq
where $Q_1$ and $Q_2$ are two different heavy quarks
of mass $m_1$ and $m_2$ respectively
and the non-$QCD$, i.e.\ non colored, partons, 
can be a photon, a lepton pair, 
an intermediate vector boson, etc.
For the decay to occur, one has to assume:
\beq
m_2 \, < \, m_1.
\eeq


Let us describe in qualitative terms the physics
of the simplest process above as far as soft-gluon
dynamics is concerned, namely the rare decay
\beq
\label{eq_def_BtoXsgamma}
B \, \to \, X_s \, + \, \gamma,
\eeq
where $X_s$ is the final hadronic state
containing the strange quark $s$, coming
from the fragmentation of the beauty quark $b$
inside the $B$ meson.
In order to construct a general theory, let us assume 
that the strange quark mass $m_s$ is a parameter that
we can change at will.
Let us consider first the massless limit of the final quark,
\beq
m_s \, = \, 0.
\eeq
We assume to be in the so-called threshold 
(or large-$x$) region,
\beq
\label{eq_thres_reg_massless}
m_{X_s}^2 \, \ll \, m_b^2, 
\eeq
in which the invariant mass of the final hadronic
(partonic in $pQCD$) state $X_s$ is restricted to be
much smaller than the hard scale of the process, provided
by the initial beauty quark mass, $Q=m_b$.
In terms of the normalized invariant mass squared
\beq
\label{eq_def_y_massless}
y \, \equiv \, \frac{m_{X_s}^2}{m_b^2}
\qquad (m_s=0),
\eeq
the threshold region is simply written:
\beq
\label{restr_y_var}
y \, \ll \, 1
\qquad (y \in [0,1]).
\eeq
Roughly speaking, in the threshold region,
not much radiation can be emitted, so
the related rate is expected to be suppressed.
Note that we only fix the invariant mass $m_{X_s}$
of the final hadronic state, and not
other quantities, such as for example
the strange quark energy or its 
transverse momentum with respect to the photon
3-momentum.
The final hadronic state $X_s$ is treated
as a single pseudo-particle, with
a continuous invariant mass distribution
(rather than a fixed mass, like an ordinary particle).
We may say that, by means of the condition 
(\ref{restr_y_var}), we observe $QCD$ radiation
indirectly, in a semi-inclusive way.

The final strange quark, assumed to be
emitted with a large energy compared to the $QCD$
scale $\Lambda_{QCD} \approx 300 \, \mathrm{MeV}$
for $pQCD$ to be relevant, 
\beq
E_s \, \approx \, \frac{m_b}{2} \, \gg \, \Lambda_{QCD},
\eeq
evolves, because of collinear emissions,
into a hadronic jet.

 
Let us now consider the rare decay (\ref{eq_def_BtoXsgamma})
in the general massive case $m_s \ne 0$.
The situation becomes substantially more complicated
because of the presence of a new mass scale.
The definition of the threshold region 
(\ref{eq_thres_reg_massless}) can
be generalized by means of the condition
\beq
\label{eq_thr_reg_massive_gen}
m_{X_s}^2 \, - \, m_s^2 \, \ll \, m_b^2 
\qquad (m_s \ne 0).
\eeq
The invariant mass $m_{X_s}$ of the final hadronic state $X_s$ 
is restricted to not become much larger than $m_s$, 
compared to the hard scale $m_b$.
As in the massless case, 
that is again a constraint on $QCD$ radiation:
the latter cannot increase too much $m_{X_s}$
with respect to  $m_s$. 
Note that the condition above trivially
reduces to (\ref{eq_thres_reg_massless}) in the massless case,
so it is a sensible generalization.
Let us also remark that the condition (\ref{eq_thr_reg_massive_gen}) does not imply neither $m_{X_s}^2 \ll m_b^2$
nor $m_s^2 \ll m_b^2$.

The unitary adimensional variable $y$ 
defined in eq.(\ref{eq_def_y_massless}) is naturally
generalized, in the massive case, as
\beq
y \, \equiv \, \frac{m_{X_s}^2 \, - \, m_s^2}{m_b^2 \, - \, m_s^2}
\qquad 
\qquad (m_s \ne 0),
\eeq
in terms of which the threshold region is written
\beq
y \, \ll \, 1,
\eeq
just like in the massless case.
We have divided the squared mass increase,  
$m_{X_s}^2 \, - \, m_s^2$,
by $m_b^2 \, - \, m_s^2$, instead of simply $m_b^2$,
in order to have a unitary variable 
$(y \in [0,1])$,
again as in the massless case.

If the final strange quark is relativistic
(in the beauty rest frame),
the non-vanishing of $m_s$ produces the
well-known dead-cone ($dc$) effect, i.e.\ the
fact that gluon radiation is mostly
emitted outside a cone centered
around the strange quark motion direction, 
namely
\beq
\label{diseq_cond1}
\theta \, \gsim \, \theta_{\mathrm{dc}} \, = \, \frac{m_s}{E_s} \, = \, \frac{1}{\gamma_s}
\, \ll \, 1.
\eeq
$\gamma_s$ is the Lorentz factor of the strange quark,
\beq  
\gamma_s \, \equiv \, 
\frac{1}{ \sqrt{1 \, - \, \left| \vec{v}_s \right|^{\,2} / c^2} } 
\, \gg \, 1,
\eeq
with $\vec{v}_s \equiv d\vec{r}_s/dt$ the ordinary strange quark 3-velocity.
The dead-cone effect is well known from classical electrodynamics
\cite{Jackson:1998nia};
In general, a non-vanishing final quark mass softens the collinear singularity according to a general mechanism
\cite{Aglietti:2006wh}.
Since the source of kinetic energy of the strange
quark is the beauty mass, the strange quark
can be relativistic only if it is much lighter than
the beauty, i.e.\ if
\beq
m_s \, \ll \, m_b. 
\eeq
Since the usual kinematic condition of $y$ 
larger than a given positive value $y_{\mathrm{cut}}$,
namely 
\beq
\label{diseq_cond2}
y \, > \, y_{\mathrm{cut}} \, > \, 0,
\eeq
also gives a constraint on gluon emission
angles (the related observable is an infrared safe quantity),
one has to find which one of the two limitations
(\ref{diseq_cond1}) and (\ref{diseq_cond2})
is stronger and then effective.


In general, we can identify three different subregions in
the threshold region (\ref{eq_thr_reg_massive_gen}).
\begin{enumerate}
\item
{\it The effectively-massless region}, 
in which the strange mass is much smaller
than the final jet mass,
\beq
\label{diseq_cond_effect_nomass}
m_s^2 \, \ll \, m_{X_s}^2 \, \ll \, m_b^2.
\eeq
In this case, the strange mass only gives power corrections
to the massless distribution previously considered $(m_s=0)$,
of the form
\beq
\left(\frac{m_s^2}{m_{X_s}^2}\right)^n,
\qquad n=1,2,3,\cdots,
\eeq
possibly multiplied by logarithms of the same quantity.
As already remarked, since the invariant mass distribution
is an infrared (i.e.\ soft and collinear) safe quantity, 
no strange quark mass singularities can arise
(i.e.\ terms of the form $\log m_s^2/m_{X_s}^2$
without a power-suppressed coefficient).
In this case, the strange mass is so small that
the dead-cone effect (\ref{diseq_cond1}) gives 
a small correction to the massless distribution.
Note that relation (\ref{diseq_cond_effect_nomass})
is basically equivalent to the relation
\beq
m_s^2 \, \ll \, m_{X_s}^2 \, - \, m_s^2 \, \ll \, 
m_b^2 \, - \, m_s^2 \approx m_b^2.
\eeq
If we define the mass-correction parameter
\beq
\rho \, \equiv \, \frac{m_s^2}{m_b^2 \, - \, m_s^2}
\qquad (0 < \rho < \infty),
\eeq
the region (\ref{diseq_cond_effect_nomass}) is simply written
\beq
\rho \, \ll \, y \, \ll \, 1.
\eeq
\item
{\it The quasi-collinear slice}, 
in which the increase of the jet invariant mass
produced by soft-gluon radiation is comparable
to the strange quark mass, 
\beq
m_{X_s}^2 \, - \, m_s^2 \, \approx \, m_s^2 \, \ll \, m_b^2,
\eeq
or, equivalently,
\beq
\rho \, \approx \, y \, \ll \, 1.
\eeq
Formally, one can consider the (correlated) limit:
\beq
y \, \to \, 0^+, 
\qquad \rho \, \to \, 0^+, 
\qquad \frac{\rho}{y} \, \to \, \mathrm{const},
\eeq
where $\mathrm{const} \ne 0, \infty$.
This is a double-logarithmic region, 
as the previous one or the massless case.
\item
{\it The soft region}, in which the increase 
in the final jet mass due to gluon radiation
is much smaller than the strange mass,
\beq
m_{X_s}^2 \, - \, m_s^2 \, \ll \, m_s^2.
\eeq
In terms of the adimensional variables
we have introduced, the above condition is written:
\beq
y \, \ll \, \rho.
\eeq
Since we always assume $y \ll 1$, the
above relation basically implies one of the two
following possibilities:
\beq
\rho \, \approx \, 1
\quad \mathrm{or} \quad \rho \, \gg \, 1.
\eeq
Since the final strange quark is not relativistic
in any of the two above cases, 
there is not any collinear enhancement in this region.
At any order of perturbation theory,
one finds at most, in the invariant mass distribution,
one large infrared logarithm
of soft origin for each power of $\alpha_S$.
In other words, this region is a single-logarithmic,
rather than a double-logarithmic, one%
\footnote{
The situation is analogous to Deep Inelastic Scattering
$(DIS)$ (even though the latter is not a collinear safe process),
where soft singularities, unlike collinear ones,
cancel in the inclusive cross section.
Therefore the perturbative expansion of the latter
contains at most one large infrared logarithm
of collinear origin for each power of $\alpha_S$
\cite{Dokshitzer:1991wu}. 
}.
If $\rho \gg 1$, the final quark is very slow
(in the initial quark rest frame) and soft-gluon radiation
is suppressed by color coherence.
Soft gluons indeed "see" a static color charge
which, at the fragmentation time, begins to move
with a very small velocity,
without any color-spin flip.
In the limit of vanishing final velocity,
soft gluons just see a static color charge at any time.

\end{enumerate}
The paper is organized as follows.
In sec.$\,$\ref{sec_phen_rel} we discuss
the main phenomenological applications
of our work. Nature provides to us
heavy quark decays with quite
different mass ratios, so we conclude that
our work is not academic.

In sec.$\,$\ref{sec_massless} we consider
threshold resummation in the massless
limit of the final quark, $m_2=0$.
As already anticipated,
this case is considerably simpler than
the massive case $m_2 \ne 0$, which is our
primary concern.
This is a preliminary section written 
in order to 
present the main ideas in a simple case
and to fix the notation. 
This section also has a pedagogical character
and can be omitted by an expert on
threshold resummation.

In sec.$\,$\ref{sec_first_order}
we describe the exact calculation
to first-order in $\alpha_S$, 
of the photon energy spectrum in the rare
$B \to X_s \gamma$ decay, 
which is assumed as a model process.
As already remarked,
the above process is selected because
of its simplicity, but we believe that the
main consequences which we derive can
be generalized to more complicated
processes in the class (\ref{general_process}),
and perhaps even more.

In sec.$\,$\ref{sec_soft_factor} 
we consider threshold resummation,
in the usual rare decay, in the soft limit.
As already remarked, that means, 
that the (massive) final quark is produced,
in the fragmentation of the initial heavy quark
(at rest), with a non-relativistic velocity.
We may say that the latter is a complementary situation
with respect to the massless limit
of the final quark.

In sec.$\,$\ref{sec_gener_factor}
we construct a general factorization
scheme in the massive case.
We introduce, as usual,
a universal, i.e. process
independent, long-distance dominated 
$QCD$ Form factor, resumming the
infrared logarithms to all orders in
$\alpha_S$, together with a Coefficient and a 
Remainder functions.
Unlike the form factor, the latter 
are process-dependent, short-distance quantities,
having an ordinary (i.e. truncated) 
perturbative expansion.

Sec.$\,$\ref{sec_improv_factor_scheme}
is the central one, the core of the paper.
In this section
we consider the problems of the
general massive factorization scheme,
constructed in the previous section,
concerning the massless limit 
of the final quark,
$m_2 \to 0$,
which turns out not to be correctly reproduced.
An improved factorization scheme
is then constructed which reproduces,
in the massless limit, the factorization
of the massless process discussed in 
sec.$\,$\ref{sec_massless}.
The main point is that,
as we are going to show, it is necessary
to introduce a dependence, inside the
Coefficient function, on the final hadron
invariant mass, i.e.\ on the variable $y$.

Finally, sec.$\,$\ref{sec_conclus} contains the 
conclusions of our analysis, together with a 
discussion about future developments.
In general, a lot of work along the lines of 
this paper remains to be made.
The main developments which we can foresee, 
involve the application
of the improved factorization scheme
to other processes than $B\to X_s\gamma$
decays, as well as the generalization
of the scheme to higher orders. 


\section{Phenomenological Relevance}
\label{sec_phen_rel}

As far as soft-gluon effects are concerned,
the hard scale $Q$ of the process (\ref{general_process}),
in the rest frame of the initial heavy quark $Q_1$
($\vec{p}_1=0$), is given by
\cite{Aglietti:2001br, Lange:2005yw, Andersen:2005mj, 
Aglietti:2005mb, Aglietti:2005bm, Aglietti:2005eq}:   
\beq
Q \, = \, E_X \, + \, \left| \vec{p}_X \right|,
\eeq
where, as already defined, $X$ denotes the final hadronic 
state into which the quark $Q_2$ evolves
(basically, a hadronic jet).
If we denote by $q^\mu$ the total 4-momentum
of the {\it non QCD partons}, the hard scale
can be written:
\beq
\label{rel_hard_scale}
Q \, = \, m_1
\, - \, \sqrt{q^2 \, + \, \left| \vec{q} \, \right|^2} 
\, + \, \left| \vec{q} \, \right|
\, = \, m_1 \, - \, q_0 \, + \, \sqrt{q_0^2 \, - \, q^2},
\eeq
where 
\beq
q^2 \, \equiv \, q^\mu q_\mu \, = \, 
q_0^2 \, - \, \left| \vec{q} \, \right|^2
\eeq
is the invariant mass squared of the {\it non QCD partons}.
Note that, for large values of $q^2$,
the non colored particles can take away a substantial
fraction of the available energy from the
$QCD$ subprocess, reducing to a large extent
the hard scale $Q$ from the "natural" or upper
value $m_1$:
\beq
q^2 \, \lsim \, m_1^2 
\quad \Rightarrow \quad
Q^2 \, \ll \, m_1^2.
\eeq
In the real world, one may cite the following
cases of heavy-to-heavy decays (\ref{general_process}):
\begin{enumerate}
\item
\label{case1}
The rare (one-loop mediated) beauty quark decays
\beq
\label{eq_rare_decay}
b \, \to \, X_s \, + \, \gamma,
\eeq
which we have already considered in the Introduction.
In the real world, if we take a constituent (i.e.\ large) strange quark mass $m_s \approx 500 \, \mathrm{MeV}$ 
(let's say, one half of the $\Phi$ mass),
the quark mass ratio
\beq
\frac{m_s}{m_b} \, \approx \, \frac{1}{10}.
\eeq 
In this process, since the photon is real, 
the 4-momentum $q^\mu$ of the probe is light-like, 
$q^2=0$,
so the hard scale $Q$, according to eq.(\ref{rel_hard_scale}),
exactly coincides with the beauty mass:
\beq
Q \, = \, m_b;
\eeq
\item
The $CKM$-favored semileptonic $b$ decays
\beq
\label{eq_semilep}
b \, \to \, X_c \, + \, l \, + \, \nu,
\eeq
where $X_c$ is the final hadronic state 
containing the charm quark, 
coming from beauty fragmentation,
with the (rather large) quark mass ratio
\beq
\frac{m_c}{m_b} \, \approx \, \frac{1}{3}.
\eeq
As already noted, the lepton pair can take away
a considerable energy from the $QCD$ subprocess.
Unlike previous case \ref{case1}, 
according to eq.(\ref{rel_hard_scale}),
the true hard scale
\beq
Q \, = \, E_{X_c} \, + \, \left| \vec{p}_{X_c} \right|,
\eeq
is substantially smaller than the beauty mass $m_b$
for a large dilepton invariant mass, 
$q^2 \lsim m_b^2$;
\item
As a final example of (\ref{general_process}),
let us mention the $CKM$-favored top quark decays
\beq
t \, \to \, b \, + \, W,
\eeq
where the heavy quark mass ratio is very small: 
\beq
\frac{m_b}{m_t} \, \approx \, \frac{1}{35}.
\eeq
\end{enumerate}
We may conclude that phenomenology offers
a rather wide class of processes (\ref{general_process}),
with quite spread values of the quark mass ratio.


\section{Massless Case}
\label{sec_massless}

For simplicity's sake, let us begin our analysis
considering the rare decay (\ref{eq_rare_decay})
in the massless limit of the final strange quark,
\beq
m_s \, = \, 0.
\eeq
In this heavy-to-light decay, 
both fixed-order and resummed calculations
greatly simplify.
Again for simplicity's sake, let us approximate the
effective weak nonleptonic Hamiltonian governing the decay 
(\ref{eq_rare_decay}) by keeping only
the local operator
\bea
O_7(x) &\equiv& \frac{e}{16 \, \pi^2} \, 
\bar{s}_i(x) \,
\sigma^{\mu\nu} \left( m_b \, R \, + \, m_s \, L \right)
b^i(x) \, F_{\mu\nu}(x)
\, = \,
\\
&=& \frac{e \, m_b}{16 \, \pi^2} \, 
\bar{s}_i(x) \,
\sigma^{\mu\nu}  R \,b^i(x) \, F_{\mu\nu}(x)
\qquad\qquad\qquad\qquad\qquad\qquad\qquad (m_s = 0),
\nonumber
\eea
where $e$ is the proton charge and
we have defined the standard Right $(R)$ 
and Left $(L)$ projectors:
\beq
R \, \equiv \, \frac{1 \, + \, \gamma_5}{2};
\qquad
L \, \equiv \, \frac{1 \, - \, \gamma_5}{2}.
\eeq
The generalization to all the operators 
in the effective Hamiltonian will be discussed
in sec.$\,$\ref{new_sec_gen}.


\subsection{Total decay rate}

The tree-level width of the rare decay (\ref{eq_rare_decay})
reads:
\beq
\label{eq_Gamma_0_r_eq_0}
\Gamma^{(0)}_0 \, = \, \frac{G_F^2 m_b^5}{32 \, \pi^3} \, 
C_7^2 \, \left| \lambda_t \right|^2
\, \frac{\alpha_{em}}{\pi}, 
\eeq
where $G_F$ is the Fermi constant, $m_b$ is the on-shell
beauty mass, $\alpha_{em} \approx 1/137$
is the fine-structure constant
and the constant $C_7$ is the Wilson (short-distance) 
Coefficient function of the operator 
$O_7$,
resumming large logarithms of $m_w/m_b$ to all orders, 
as well as collecting finite corrections. 
Finally $\lambda_t$ is a product of $CKM$ matrix
elements: 
\beq
\lambda_t \, \equiv \, V_{tb} V_{ts}^*,
\eeq
Through the paper, we use the following conventions:
the lower index zero, on the l.h.s.\ of eq.(\ref{eq_Gamma_0_r_eq_0})
for example, refers to the massless limit.
In general, we will denote quantities calculated in the
$m_s=0$ limit with a zero subscript.
The upper index, between round brackets, denotes
instead the order in perturbation theory.


\subsection{Photon spectrum or invariant hadron squared mass  distribution}

We consider the invariant hadron squared-mass spectrum,
paying particular attention to the low-mass
or threshold region
\beq
\label{eq_low_mass}
m_{X_s}^2 \, \ll \, m_b^2.
\eeq
As already noted, the hard scale $Q$ is given by the 
beauty mass $m_b$, as:
\beq
Q \, = \, E_{X_s} \, + \, |\vec{p}_{X_s}| \, = \, 
m_b \, - \, E_\gamma \, + \, |\vec{p}_\gamma| \, = \, m_b
\qquad (c=1).
\eeq
Since, at lowest order in the $QCD$ coupling $\alpha_S$,
there is no gluon radiation,
the final hadronic state only contains the
strange quark,
\beq
\alpha_S \, = \, 0: \quad X_s \, = \, s,
\eeq
so that:
\beq
m_{X_s}^ 2 \, = \, m_s^2 \, = \, 0.
\eeq
By defining the unitary variable
\beq
y \, \equiv \, \frac{m_{X_s}^2 \, - \, m_s^2}{m_b^2 \, - \, m_s^2}
\, = \, \frac{m_{X_s}^2}{m_b^2}
\qquad  (m_s = 0),
\eeq
it follows that the tree-level (i.e.\ lowest-order) spectrum is a spike
at vanishing $y$:
\beq
\label{tree_lev_spec}
\frac{d\Gamma_0^{(0)}}{dy} \, = \, \Gamma^{(0)}_0 \, \delta(y).
\eeq
In general, the differential spectrum in $y$ has a perturbative expansion
in powers of $\alpha_S$ of the form:
\beq
\frac{d\Gamma_0}{dy}\left( y; \alpha_S \right)
\, = \, \Gamma_0^{(0)} \, \delta(y) \, + \, \sum_{n=1}^\infty 
C_F \left( \frac{\alpha_S}{\pi} \right)^n  
\frac{d\Gamma_0^{(n)}}{dy}(y)
\qquad (0 \le y \le 1),
\eeq
where $C_F = (N^2-1)/(2N)=4/3$ for $N=3$ colors in $QCD$.
In the beauty rest frame ($p_b^\mu = (m_b;0,0,0)$), 
by elementary kinematics:
\beq
E_\gamma \, = \, \frac{m_b^2 \, - \, m_{X_s}^2}{2 m_b}
\, = \, \frac{m_b}{2} 
\left( 1 \, - \, \frac{m_{X_s}^2}{m_b^2} \right).
\eeq
By defining the unitary variable
\beq
x \, \equiv \, \frac{E_\gamma}{\, E_\gamma^{\max}} 
\, = \, 1 \, - \, \frac{m_{X_s}^2}{m_b^2}
\qquad (0\le x \le 1),
\eeq
we find the relation
\beq
x \, = \, 1 \, - \, y \qquad (m_s=0).
\eeq
Therefore, to evaluate the hadron squared mass distribution
is equivalent to compute the photon energy spectrum.
In the high-mass region,
the emitted photon is soft,
while in the low-mass region (\ref{eq_low_mass}), 
the photon is hard,
i.e.\ its energy $E_\gamma$ is close to its upper endpoint $m_b/2$,
so that:
\beq
E_\gamma^{\min} \, = \, 0;
\qquad  
E_\gamma^{\max} \, = \, \frac{m_b}{2}
\qquad (m_s=0).
\eeq
For technical reasons (to avoid distributions), it is simpler to consider
the normalized, partially-integrated
spectrum, the so-called event fraction:
\beq
\label{eq_define_EF}
E_0(y;\alpha_S) \, \equiv \,
\frac{1}{\Gamma_0} 
\int\limits_0^y \frac{d\Gamma_0}{dy'}(y';\alpha_S) \, dy'
\qquad (0 \le y \le 1). 
\eeq
The differential spectrum is simply obtained
by differentiation of the event fraction:
\beq
\frac{1}{\Gamma_0} 
\frac{d\Gamma_0}{dy}
\, = \, 
\frac{dE_0}{dy}.
\eeq
It follows directly from the definition 
of event fraction given in eq.(\ref{eq_define_EF}) 
that:
\beq
\label{eq_R_constraints}
\lim_{y \to 0^+} E_0(y;\alpha_S) \, = \, 0;
\qquad
\lim_{y \to 1^-} E_0(y;\alpha_S) \, = \, 1.
\eeq
By integrating both sides of eq.(\ref{tree_lev_spec})
with respect to $y$, it is immediately found that
the tree-level event fraction is identically equal to one
for any $y>0$:
\beq
E_0^{(0)}(y) \, \equiv \, \frac{1}{\Gamma_0^{(0)}} \int\limits_0^y
\frac{d \Gamma_0^{(0)}}{dy'} \, dy' \, = \, \theta(y) \, \equiv \, 1 \qquad (y>0);
\eeq
where $\theta(y) \equiv 1$ for $y>0$
and zero otherwise is the standard Heaviside unit-step
function.

The event fraction at first order in $\alpha_S$ only depends
on diagrams involving single real gluon emission
(bremmstrahlung), as:
\beq
\label{R_first_ord_brem}
E_0(y;a) \, = \, \frac{\Gamma_0^{(0)}
\, + \,
a \, \int\limits_0^y dy' \, d\Gamma_0^{(1)}/dy' 
\, + \, \mathcal{O}\left( a^2 \right)
}{
\Gamma_0^{(0)} \, + \,  a \, 
\int\limits_0^1 dy \, d\Gamma_0^{(1)}/dy 
\, + \, \mathcal{O}\left( a^2 \right)
}
\, = \, 1 \, - \, \frac{a}{\Gamma_0^{(0)}} \,
\int\limits_y^1  \frac{d\Gamma_0^{(1)}}{dy'} \, dy'
\, + \, \mathcal{O}\left( a^2 \right)
\qquad (y>0);
\eeq
where we have defined the effective, first-order coupling
of the (heavy) quarks to gluons
\beq
a \, \equiv \, \frac{C_F \alpha_S }{\pi}.
\eeq
We may say that the event fraction and the total rate give complementary
information about the decay process.
Note that the second equation in (\ref{eq_R_constraints})
is trivially satisfied by eq.(\ref{R_first_ord_brem}).
To verify the first equation is instead less trivial.
To accomplish this task,
one has to resum soft-gluon effects to all
orders in $\alpha_S$, as we are going to show.

\begin{figure}[ht]
\begin{center}
\includegraphics[width=0.5\textwidth]{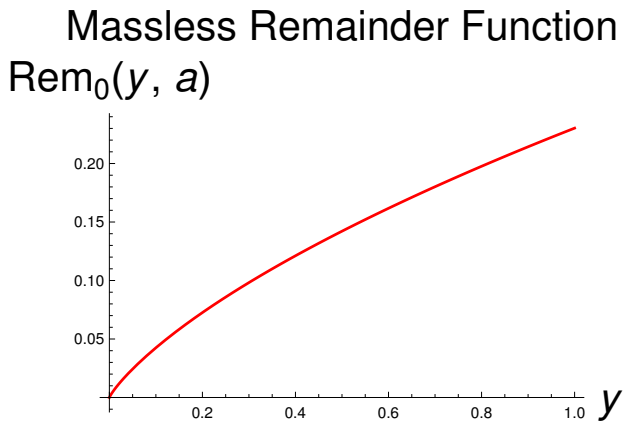}
\footnotesize
\caption{
\label{fig_PlotRemD}
\it
Plot of Massless Remainder Function $\mathrm{Rem}_0(y;a)$ 
in first approximation, 
eq.(\ref{eq_Rem_Funct_Massless}),
as a function of $y$ in all its kinematic range,
$0 \le y \le 1$.
We have defined $a \equiv C_F \alpha_S(m_b) / \pi \cong 0.089$
for $\alpha_S(m_b)=0.21$.
}
\end{center}
\end{figure}

\noindent
An exact first-order calculation in $\alpha_S$
or, equivalently, in $a$, 
of the event fraction gives \cite{Ali:1990tj,Ali:1995bi}:
\bea
E_0\left( y; a \right) \, = \, 1 
&-&  \frac{a}{2} \, \log^2(y)
\, - \, \frac{7}{4} \, a  \, \log(y)
\, - \, a \, \frac{31}{12} \, + 
\\
&-& a
\bigg[
\, y \left( 1 \, - \, \frac{y}{4} \right) \log(y)
\, + \, \frac{y}{12} \left( 2 \, y^2 \, - \, 3 \, y \, - \, 30 \right)
\bigg]
\, + \, \mathcal{O}\left( a^2 \right).
\nonumber
\eea
The spectrum above contains three different kind of terms,
as far as the Born-kinematics limit $y \to 0^+$ 
is concerned:
\begin{enumerate}
\item
{\it Double and single logarithmic terms of $y$},
namely the terms
\beq
- \, \frac{a}{2} \, \log^2(y), 
\qquad
\, - \, \frac{7}{4} \, a \, \log(y),
\eeq
which formally diverge in the limit $y \to 0^+$,
and which are therefore very large in the
lower end-point region $y \ll 1$ --- namely the threshold region.
These are clearly the dominant terms in the small-$y$ region;
\item
{\it Constant terms with respect to $y$}, 
namely the term
\beq
\, - \,
a \, \frac{31}{12}.
\eeq
In units of the ubiquitous factor $a$, 
the constant
\beq
\label{eq_give_C1}
C^{(1)}_0 \, = \, - \, \frac{31}{12} \, \cong \, - \, 2.6
\eeq
is of order one, as expected.
For $\alpha_S(m_b)=0.21$, the first-order correction turns
out to be 
\beq
a \, C^{(1)}_0 \, \simeq \, - \, 0.23;
\eeq
\item
{\it Infinitesimal terms in $y$,} i.e.\ terms
which vanish in the limit $y \to 0^+$,
namely
\beq
a \, H^{(1)}_0(y),
\eeq
where:
\beq
\label{eq_H1}
H^{(1)}_0(y) 
\, \equiv \,
- \, y \left( 1 \, - \, \frac{y}{4} \right) \log(y)
\, - \, \frac{y}{12} \left( 2 \, y^2 \, - \, 3 \, y \, - \, 30 \right).
\eeq
These latter terms are the least important ones
in the small-$y$ region, but give a substantial
contribution in the bulk of the spectrum, 
i.e.\ for $y = \mathcal{O}(1)$.
These terms can be neglected, to a first approximation,
in the small-$y$ region, but
cannot be neglected anymore for generic $y$ values,
where they are not smaller than the logarithmic
or the constant terms.
\end{enumerate}
As far as the small-$y$ behavior is concerned,
in the general process (\ref{general_process}),
the event fraction is naturally written 
to first order in the form 
\beq
\label{eq_exact_first_ord}
E_0(y; a) \, = \, 1 
\, - \, \frac{a}{2} \, A^{(1)} \log^2(y) 
\, + \, a \, S^{(1)} \log(y)
\, + \, a \, C^{(1)}_0 
\, + \, a \, \mathrm{Rem}^{(1)}_0(y) 
\, + \, \mathcal{O}\left( a^2 \right);
\eeq 
where we have introduced the coefficients:
\beq
\label{eq_give_A1_and_S1}
A^{(1)} \, = \, 1;
\qquad 
S^{(1)} \, = \, - \, \frac{7}{4}.
\eeq
The value of the constant $C^{(1)}_0$ has
already been given in eq.(\ref{eq_give_C1})
and the first-order contribution to the remainder
function
\beq
\mathrm{Rem}^{(1)}_0(y) \, \equiv \, H^{(1)}_0(y)
\, \to \, 0
\,\,\, \mathrm{for} \,\, y \, \to \, 0^+.
\eeq
The complete Remainder function at first order,
\beq
\label{eq_Rem_Funct_Massless}
\mathrm{Rem}_0(y;a) \, = \,
a \, \mathrm{Rem}^{(1)}_0(y), 
\eeq
is plotted in fig.$\,$\ref{fig_PlotRemD}.
Note that, since $E_0(y=1,a) \equiv 1$,
it follows that:
\beq
\mathrm{Rem}^{(1)}_0(y=1) \, = \, - \, C^{(1)}_0 \, = \, \frac{31}{12}.
\eeq
Actually, the Remainder Function is 
a strictly-monotonically-increasing function of $y$,
and therefore positive in all its range,
$0 \le y \le 1$.
%
%
%
%

The constants $A^{(1)}$ and $S^{(1)}$ are process-independent,
i.e.\ they are the same for all the heavy-to-light decays in the 
class (\ref{general_process}).
That is a consequence of the general properties
of $QCD$ radiation in the infrared (i.e.\ soft and/or collinear) 
limit.
On the contrary, the Coefficient $C^{(1)}_0$ and the Remainder
function $\mathrm{Rem}^{(1)}_0(y)$
are short-distance dominated and are therefore process dependent.

In the simple case of the radiative decay (\ref{eq_rare_decay})
-- a two-body decay at tree level --
$C^{(1)}_0$ is truly a constant and $\mathrm{Rem}^{(1)}_0(y)$ only depends
on $y$.
In more complicated heavy-to-light decays, such as for
example the semileptonic $b \to u l \nu$ decays
-- 3-body decays at tree-level -- 
$C^{(1)}$ still does not depend on $y$, but
it does depend on other kinematical 
variables.
Similarly, the Remainder function $\mathrm{Rem}^{(1)}(y)$ also
depends on additional kinematical 
variables.


\subsection{Factorization}

The basic idea of factorization 
is simply to separate from each other perturbative terms
having different physical origin.
In particular, we factorize the large infrared logarithms
into a universal, i.e. process independent, form factor.
To order $a$ (i.e.\ to order $\alpha_S$), one can write indeed:
\beq
\label{eq_resum_first_ord}
E_0(y;a) \, = \, C_0(a) \, \Sigma_0(y; a)
\, + \, \mathrm{Rem}_0(y; a) 
\, + \, \mathcal{O}\left( a^2 \right);
\eeq 
where we have defined the long-distance dominated
$QCD$ form factor
\beq
\label{QCD_FF_massless_ord1}
\Sigma_0(y;a) \, \equiv \, 
1 \, - \, \frac{a}{2} \, A^{(1)} \log^2(y) 
\, + \, a \, S^{(1)} \log(y);
\eeq
the short-distance Coefficient function:
\beq
C_0(a) \, \equiv \, 1 \, + \, a \, C^{(1)}_0
\eeq
and the short-distance Remainder function
\beq
\mathrm{Rem}_0(y;a) \, \equiv \, a \, H^{(1)}_0(y).
\eeq
Note that the perturbative expansion of the Remainder function
begins at order $a$, i.e.\ it vanishes in the
free limit $a \to 0$,
while the Form factor and the Coefficient function 
equal unity in the same limit.

By expanding the product on the r.h.s.\ of eq.(\ref{eq_resum_first_ord})
in powers of $a$, one finds: 
\beq
\label{eq_factor_expand}
E_0(y;a) \, = \, 1  
\, + \, a \, \Sigma_0^{(1)}(y)
\, + \, a \, C^{(1)}_0
\, + \, a \, \mathrm{Rem}^{(1)}_0(y) 
\, + \, \mathcal{O}\left(a^2\right).
\eeq 
On the r.h.s.\ of the above equation, one finds
exactly the same
first-order terms which are on the r.h.s.\ of the
fixed-order expansion, eq.(\ref{eq_exact_first_ord}).
Therefore factorization, 
which we have explicitly constructed at first order
in $a$, 
involves a shift of terms of second 
(and in general also higher) order.

Let us note that we have constructed a {\it minimal scheme}
for the $QCD$ form factor $\Sigma_0(y;a)$, inside which only
logarithmic terms of $y$ are included.
In other words, constants and infinitesimal terms
for $y \to 0^+$ are not included in our $\Sigma_0$.
Let us also remark that the factorization
scheme given by eq.(\ref{eq_resum_first_ord})
can be consistently pushed to higher orders in $\alpha_S$.


\subsection{Threshold resummation in the heavy-to-light case}
\label{sec_TR_HtoL}

The $QCD$ form factor, resumming to all orders in $\alpha_S$
the infrared logarithmically-enhanced terms, 
has the standard expression in moment space or $N$-space
\cite{Kodaira:1982cr,Sterman:1986aj,Catani:1989ne,Aglietti:2001br}:
\bea
\label{QCD_FF_standard}
\sigma_{0,N}\left(\alpha_S\right) &=&
\exp
\int\limits_0^1 \frac{dy}{y}
\Big[
(1-y)^{N-1} \, - \, 1
\Big]
\Bigg\{ \,\,
\int\limits_{Q^2 y^2}^{Q^2 y}
\frac{dk_\perp^2}{k_\perp^2} \,
A\left[ \alpha_S\left( k_\perp^2 \right) \right]
\, +  
\\
&& 
\qquad\qquad\qquad\qquad\qquad\qquad
\, + \, B\left[ \alpha_S\left( Q^2 y \right) \right] 
\, + \, D\left[ \alpha_S \left( Q^2y^2 \right) \right]
\Bigg\}.
\nonumber
\eea
The function $A(\alpha_S)$, having an ordinary 
perturbative expansion in powers of $\alpha_S$,
\beq
A(\alpha_S) \, = \, 
\sum_{n=1}^\infty 
C_F \left( \frac{\alpha_S}{\pi} \right)^n \, A^{(n)},
\eeq
describes soft-gluon emission at small angle
with respect to the parent strange quark 
\cite{Korchemsky:1987wg}
(i.e.\ both soft and collinear enhanced).
The first-order coefficient $A^{(1)}$
-- the only one we are directly interested to --
has been given in the first of eqs.(\ref{eq_give_A1_and_S1}),
in units of $a \equiv C_F \alpha_S/\pi$, 
according to our current conventions.

\noindent
The function $B(\alpha_S)$, also having an ordinary 
perturbative expansion in powers of $\alpha_S$,
\beq
B(\alpha_S) \, = \, 
\sum_{n=1}^\infty 
C_F \left( \frac{\alpha_S}{\pi} \right)^n \, B^{(n)}
\, = \, a \, B^{(1)} \, + \,
\mathcal{O}\left(a^2\right),
\eeq
describes hard-gluon emission at small angle
(collinear enhanced but not soft enhanced radiation).
The first-order coefficient explicitly reads:
\beq
B^{(1)} \, = \,  - \, \frac{3}{4}.
\eeq

\noindent
Finally, the function $D(\alpha_S)$, also having an ordinary 
perturbative expansion in powers of $\alpha_S$,
\beq
D(\alpha_S) \, = \, 
\sum_{n=1}^\infty 
C_F \left( \frac{\alpha_S}{\pi} \right)^n \, D^{(n)}
\, = \, a \, D^{(1)} \, + \,
\mathcal{O}\left(a^2\right),
\eeq
describes soft-gluon emission at large angle
(soft enhanced but not collinear enhanced radiation).
In heavy-to-light transitions,
the first-order coefficient takes the value:
\beq
D^{(1)} \, = \,  - \, 1.
\eeq

\noindent
It is natural to resum infrared logarithms 
in $N$-space, where, unlike physical space, 
factorization of kinematic 
constraints for multiple soft-gluon emission 
holds true \cite{Catani:1997vp}.
In $N$-space, the logarithm-enhanced terms
have the form
\beq
\alpha_S^k \, \log^l(N),
\qquad k \, = \, 1, \, 2, \, 3, \, \cdots, 
\quad 
1 \, \le \, l \, \le \, 2 \, k.
\eeq
As well known, in order to obtain the form factor in physical space
($y$-space), one has to make first analytic continuation
of $\sigma_N$ in the $N$ variable, from integer to complex
values:
\beq
N \in \NN 
\quad \mapsto \quad
\mathcal{N} \in \CC.
\eeq
The form factor in physical space
is then obtained by means of an inverse
Mellin transform:
\beq
\sigma_0(y;\alpha_S)
\, = \,
\int\limits_{c-i\infty}^{c+i\infty}
\frac{d\mathcal{N}}{2\pi i} \,
(1-y)^{-\mathcal{N}} \sigma_{0,\mathcal{N}}(\alpha_S),
\eeq
where the (real) constant $c$ is chosen
in such a way that all the singularities of 
$\sigma_\mathcal{N}$ lie to the left of the integration contour
(a vertical line in the complex $\mathcal{N}$-plane).

The partially-integrated form factor $\Sigma_0(y)$ is finally
obtained by integrating over $y$:
\beq
\label{eq_from_sig_to_Sig}
\Sigma_0(y;\alpha_S) \, = \, \int\limits_0^y 
\sigma_0(y';\alpha_S) \, dy'.
\eeq


\subsubsection{Form-factor expansion}

In order to determine the first-order Coefficient
and Remainder functions,
one has to subtract, from the event fraction,
the expansion of the form factor
up to first order in $\alpha_S$.

The first step involves expanding
the exponential on the r.h.s.\ of
eq.(\ref{QCD_FF_standard}) to first order,
\beq
\sigma_{0,N} \, = \, \exp(X)
\, = \, 1 \, + \, X \, + \, 
\mathcal{O}\left(X^2\right),
\eeq
followed by a truncation of the
resummation functions $A(\alpha_S),B(\alpha_S)$ and $D(\alpha_S)$ 
to first order in $\alpha_S$:
\bea
\label{QCD_FF_standard_esp1}
\sigma_{0,N}(\alpha_S) &=&
1 \, + \, 
\int\limits_0^1 \frac{dy}{y}
\big[
(1-y)^{N-1} \, - \, 1
\big]
\Bigg\{ \,\,
\int\limits_{Q^2 y^2}^{Q^2 y}
\frac{dk_\perp^2}{k_\perp^2} \,
A\left[ \alpha_S\left( k_\perp^2 \right) \right]
\, +  
\\
&& 
\qquad\qquad\qquad\qquad\qquad
\, + \, B\left[ \alpha_S\left( Q^2 y \right) \right] 
\, + \, D\left[ \alpha_S \left( Q^2 y^2 \right) \right]
\Bigg\}
\, + \, \mathcal{O}\left( \alpha_S^2 \right) \, =
\nonumber\\
&=&  1 \, + \, a \,
\int\limits_0^1 dy
\big[
(1-y)^{N-1} \, - \, 1
\big]
\Bigg\{ \,\,
\frac{A^{(1)}}{y} \int\limits_{Q^2 y^2}^{Q^2 y}
\frac{dk_\perp^2}{k_\perp^2}  
\, + \, \frac{ B^{(1)} \, + \, D^{(1)} }{y} 
\Bigg\}
\, + \, \mathcal{O}\left( a^2 \right);
\nonumber
\eea
where:
\beq
\alpha_S \, \equiv \, \alpha_S\left(Q^2\right);
\qquad
a \, \equiv \, \frac{C_F \, \alpha_S\left(Q^2\right)}{\pi}.
\eeq
Since $X=\mathcal{O}\left(\alpha_S\right)$,
the higher-order terms $X^2,X^3,\cdots$ 
in the expansion of $\exp(X)$ are 
of second or higher order in $\alpha_S$.
The evaluation of the inverse Mellin transform
simply gives back, at first order, the curly bracket
in the last member of eq.(\ref{QCD_FF_standard_esp1}), so that:
\beq
\sigma_0(y;\alpha_S)
\, = \, \delta(y)
\, - \, a \, A^{(1)} 
\left[ \frac{\log(y)}{y} \right]_+
\, + \, a \, \left( B^{(1)} \, + \, D^{(1)} \right) 
\left[ \frac{1}{y} \right]_+ 
\, + \, \mathcal{O}(a^2);
\eeq
where the plus regularization of a generic
function $f(y)$ is defined as
the following (weak) limit:
\beq
\left[ f(y) \right]_+
\, \equiv \,
\lim_{\epsilon\to 0^+}
\left[
\theta(y - \epsilon) \, f(y)
\, - \, \delta(y-\epsilon)
\int\limits_\epsilon^1 f(y') \, dy'
\right].
\eeq
The plus regularization comes from virtual
diagrams, related to the term $-1$
in the function $(1-y)^{N-1}-1$.
Finally, the partially-integrated form factor 
$\Sigma_0(y,\alpha_S)$,
entering the factorization formula in eq.(\ref{eq_resum_first_ord}),
is obtained by integrating over $y$
the differential form factor $\sigma_0(y,\alpha_S)$, 
according to eq.(\ref{eq_from_sig_to_Sig}):
\beq
\label{eq_QCD_FF_massless_first_ord}
\Sigma_0(y,a) \, \equiv \,
\int\limits_0^y \sigma_0(y',a) \, dy'
\, = \, 1 \, - \, \frac{ A^{(1)} }{2} \, a \, \log^2(y)
\, + \, a \, \left( B^{(1)} \, + \, D^{(1)} \right) \log(y)
\, + \, \mathcal{O}\left(a^2\right).
\eeq
The form factor above is in complete agreement
with that one in eq.(\ref{QCD_FF_massless_ord1})
if we make the identification
\beq
S^{(1)} \, = \, B^{(1)} \, + \, D^{(1)}.
\eeq
That is to say that
$S^{(1)}$ --- the coefficient of the single infrared logarithm 
at $\mathcal{O}(\alpha_S)$ ---
is the sum of the first-order collinear $B^{(1)}$
and soft $D^{(1)}$ coefficients.
The latter "separate" from each other at higher orders 
in $\alpha_S$, from second order on,
 because of the different argument of the coupling,
namely the collinear scale $Q^2 y$ 
for the function $B(\alpha_S)$ and 
the (typically much smaller) soft scale 
$Q^2 y^2$ for the function $D(\alpha_S)$,
see eq.(\ref{QCD_FF_standard}).


\section{First-order calculation in the massive case}
\label{sec_first_order}

In this section we consider an exact first-order calculation
$(\mathcal{O}(\alpha_S))$
of the photon spectrum in the rare decay (\ref{eq_rare_decay})
in the massive case $m_s \ne 0$. 


\subsection{Total rate}

By taking into account strange quark mass effects,
the tree-level width reads:
\beq
\Gamma^{(0)}(r) \, = \, \Gamma_0^{(0)} \, (1-r)^3 (1+r)
\, = \, \Gamma_0^{(0)} \, \frac{1+2\rho}{(1+\rho)^4};
\eeq
where we have defined the final-quark mass 
correction parameters
\beq
r \, \equiv \, \frac{m_s^2}{m_b^2}
\qquad (0 \le r \le 1)
\eeq
and
\beq
\rho \, \equiv \, \frac{m_s^2}{m_b^2 \, - \, m_s^2} \, = \, 
\frac{r}{1 \, - \, r}
\qquad (0 \le \rho < \infty).
\eeq
The inverse formula of the above one reads:
\beq
r \, = \, \frac{\rho}{1 \, + \, \rho}.
\eeq
The two parameters are basically the same
for small values of the strange quark mass, $m_s \ll m_b$:
\beq
\rho \, = \, r \, + \, \mathcal{O}\left(r^2\right).
\eeq
The lowest-order width in the massless limit,
$\Gamma_0^{(0)} = \Gamma^{(0)}(r=0)$, has been given in 
eq.(\ref{eq_Gamma_0_r_eq_0}).
The correction to the inclusive width at one loop 
in the massive case is given by:
\beq
\Gamma \, = \, \Gamma^{(0)}
\left[ 
1 \, + \, a \, K(\rho) 
\, + \, \mathcal{O}\left( \alpha_S^2 \right)
\right],
\eeq
where:
\bea
\label{eq_Kfactor_btosdecay}
K(\rho) &=& - \, \big(1 \, + \, 2 \, \rho \big) 
\left[
\log (\rho) \log (1 + \rho) \, + \,
2 \, \text{Li}_2(-\rho)
\, + \,  \frac{\pi^2}{3}  
\right] \, +
\nonumber\\
&& 
+ \, \rho \left(2 \, \rho^2 \, + \, 3 \, \rho \, + \, \frac{7}{2} \right) 
\Big[   
\log(1+\rho) \, - \, \log(\rho)
\Big] \, +
\nonumber\\
&& + \, 2 \, \log(1 + \rho) \, - \, 2 \, \rho \, (1 \, + \, \rho) 
\, + \, \frac{4}{3}.
\eea
The function $\mathrm{Li}_2$ is the standard dilogarithm
or Spence function:
\beq
\mathrm{Li}_2(s) \, \equiv \, - \,
\int\limits_0^s \frac{dt}{t} \log(1-t)
\, = \, \sum_{n=1}^\infty \frac{s^n}{n^2}
\qquad (|s| \, < \, 1).
\eeq
As well known, the inclusive width is an infrared safe
quantity, so that its massless limit is finite:
\beq
\lim_{\rho \to 0^+} K(\rho) \, = \, \frac{4 - \pi^2}{3}.
\eeq
Note that the most singular terms in $K(\rho)$ 
for $\rho \to 0^+$ are
of the form $\rho \log(\rho)$.


\subsection{Photon spectrum}

While, in the massive case, the lowest photon energy is still zero,
the maximal photon energy is reduced by a factor $(1-r)$
with respect to the massless case $r=0$:
\beq
E_\gamma^{\max} \, = \, \frac{m_b}{2}(1 \, - \, r).
\eeq
We consider the Event fraction $(E)$ in the kinematic variable
\beq
y \, \equiv \, \frac{m_{X_s}^2 \, - \, m_s^2}{m_b^2 \, - \, m_s^2}.
\eeq
In terms of the normalized photon energy
\beq
x \, = \, \frac{E_\gamma}{E_\gamma^{\max}}
\, = \, \frac{m_b^2-m_{X_s}^2}{m_b^2-m_s^2},
\eeq
we find the same relation of the massless case:
\beq
x \, = \, 1 \, - \, y \qquad (m_s \ne 0).
\eeq
The event fraction is naturally written:
\beq
E(y; \rho; a) \, \equiv \, 
\frac{1}{\Gamma} \int\limits_0^y
\frac{d \Gamma}{dy'}\left(y'; \rho; a\right) \, dy'
\, = \, 1 \, + \, a \, E^{(1)}(y;\rho)
\, + \, \mathcal{O}\left( a^2 \right).
\eeq
The first-order term, from the computation in \cite{Ali:1990tj},
reads in our notation:
\bea
\label{comp_Ali_and_Greub}
E^{(1)}(y;\rho) &=& + \, \big( 1 \, + \, 2 \, \rho \big)
\bigg[
- \, \log\left( \frac{y \, + \, \rho}{1 \, + \, \rho} \right) \log(y) 
\, + \, \frac{1}{2} \log^2\left( \frac{y \, + \, \rho}{1 \, + \, \rho} \right) 
\, +
\nonumber\\
&& \quad
\, + \, \log(1 \, + \, \rho) \, 
\log\left( \frac{y \, + \, \rho}{1 \, + \, \rho} \right)
\, + \, \mathrm{Li}_2\left( \frac{\rho}{y \, + \, \rho} \right)
\, - \, \mathrm{Li}_2\left( \frac{\rho}{1 \, + \, \rho} \right)
\bigg] \, +
\nonumber\\
&& - \, 2 \, \log(y) 
\, + \, \frac{1}{4} \, 
\log\left( \frac{y \, + \, \rho}{1 \, + \, \rho} \right) \, +
\nonumber\\
&& +  
\left[ 
\rho \left( 2 \, \rho^2 \, + \, 3 \, \rho \, + \, \frac{1}{2} \right)
\, - \, y \left( 1 \, - \, \frac{y}{4} \right) 
\right] 
\log\left( \frac{y \, + \, \rho}{1 \, + \, \rho} \right) \, +
\\
&& + \,
 \frac{1 \, - \, y}{y \, + \, \rho} \,
\Bigg\{
\rho \left[
2 \, \rho^2 \, + \, \rho \, (y \, + \, 2)
\, - \, \frac{1}{6}  \left(2 \, y^2 \, - \, 7 \, y \, + \, 14 \right)
\right] \, +
\nonumber\\
&& \qquad\qquad\qquad\qquad\qquad\qquad\qquad\quad
+ \, \frac{y}{12} \left(2 \, y^2 \, - \, y \, - \, 31 \right)
\Bigg\}.
\nonumber
\eea
One may notice the frequent occurrence
on the r.h.s.\ of eq.(\ref{comp_Ali_and_Greub})
of the "collinear" variable
\beq
z \, \equiv \, \frac{m_{X_s}^2}{m_b^2} 
\, = \, 
\frac{y \, + \, \rho}{1 \, + \, \rho}
\qquad \left( \frac{\rho}{1 + \rho} \le z \le 1 \right).
\eeq
Note that this dependent variable $z=z(y;\rho)$
does not vanish on Born kinematics:
\beq
\lim_{y \to 0^+} z(y;\rho) \, = \, \frac{\rho}{1 \, + \, \rho},
\eeq
but it becomes small for small $\rho$.
Note also that $z=z(y;\rho)$ exactly reduces to $y$
in the massless limit:
\beq
z(y;\rho=0) \, = \, y.
\eeq


\section{Factorization in the Soft limit}
\label{sec_soft_factor}

In this section we consider the event fraction
$E(y)$ for the rare decay (\ref{eq_rare_decay}) 
in the threshold region $y \ll 1$,
with a mass correction parameter $\rho$ of order one:
\beq
0 \, < \, y \, \ll \, \rho \, = \, \mathcal{O}(1).
\eeq
Formally, that is equivalent to taking the limit
\beq
\frac{y}{\rho} \, \to \, 0^+.
\eeq
Since the final strange quark is not relativistic
(in beauty rest frame),
there are no collinearly-enhanced terms,
so the threshold region is dominated by soft-gluon emission only
$(E_g \ll m_b)$.
As already remarked, the soft region is a single-logarithmic one, 
i.e.\ the perturbative expansion of the event fraction
$E=E(y;\rho,\alpha_S)$
contains at most one logarithm of $y$ for each
power of the coupling $\alpha_S$.


\subsection{Soft $QCD$ form factor}

The $QCD$ form factor resumming, to all orders in $\alpha_S$, 
the soft logarithmically-enhanced terms in the perturbative series
of the event fraction, 
i.e.\ in the soft limit, reads in moment space or 
$N$-space \cite{Aglietti:2007bp}%
\footnote{
We {\it argue} that the argument of $\alpha_S$ inside the function 
$\Delta(\alpha_s)$ is equal to the soft scale
$Q^2 y^2/(1 + \rho)$ --- also appearing in the coupling
inside the function $D(\alpha_s)$ ---, rather
than the collinear scale $Q^2 y^2/\rho$,
as stated in Ref.\,\cite{Aglietti:2007bp}, eq.(\ref{QCD_FF_soft}).
That is because the $D(\alpha_s)$ and $\Delta(\alpha_s)$ 
terms describe
qualitatively similar effects 
(see also eq.(\ref{QCD_FF_gen})).
However, this difference in the arguments
of the $QCD$ coupling only shows
up at $\mathcal{O}(\alpha_S^2)$, so the problem
can be definitively solved only by means of a (massive)
two-loop computation.
In any case, this problem is immaterial for the
present work, in which only $\mathcal{O}(\alpha_S)$
Coefficient and Remainder functions are evaluated.
}:
\bea
\label{QCD_FF_soft}
\sigma_N^{(S)}\left(\rho,\alpha_S\right) &=&
\exp
\int\limits_0^1 \frac{dy}{y}
\Big[
(1-y)^{N-1} \, - \, 1
\Big]
\Bigg\{ \,\,
\int\limits_{Q^2 y^2/(1 + \rho)}^{Q^2 y^2/\rho}
\frac{dk_\perp^2}{k_\perp^2} \,
A\left[ \rho ; \alpha_S\left(k_\perp^2\right) \right]
\, +  
\\
&& \qquad\qquad\qquad\qquad\qquad\qquad
+ \, D\bigg[ \alpha_S \Big( \frac{Q^2y^2}{1+\rho} \Big) \bigg]
\, + \, \Delta\bigg[\alpha_S\Big( \frac{Q^2y^2}{1 + \rho} \Big) \bigg] 
\Bigg\}.
\nonumber
\eea
The function $A(\rho;\alpha_S)$
is a "massive", i.e.\ $\rho$-dependent,
generalization of the usual massless
function $A(\alpha_S)$, reducing to the latter
in the massless limit:
\beq
\lim_{\rho \to 0^+} A(\rho;\alpha_S) \, = \, A(\alpha_S).
\eeq
The first-order term reads:
\beq
A^{(1)}(\rho) \, = \, 1 \, + \, 2 \, \rho.
\eeq
The function 
\beq
\Delta(\alpha_S) \, = \, \sum_{n=1}^\infty C_F
\left( \frac{\alpha_S}{\pi} \right)^n \Delta^{(n)}
\, = \, a \, \Delta^{(1)} \, + \, \mathcal{O}\left(a^2\right)
\eeq
describes soft parton emission off the (massive) strange quark line.
The first-order term takes the value:
\beq
\Delta^{(1)} \, = \, - \, 1.
\eeq


\subsubsection{Form factor expansion}

By expanding the resummed form factor in eq.(\ref{QCD_FF_soft})
to first order in $\alpha_S$, as described in sec.$\,$\ref{sec_TR_HtoL},
one obtains for the partially-integrated form factor
in physical space:
\beq
\Sigma_S\left(y; \rho; a \right) 
\, \equiv \,
\int\limits_0^y \sigma_S\left(y'; \rho; a \right) \, dy' 
\, = \,
1 \, + \, a \, \Sigma_S^{(1)}(y;\rho)
\, + \,\mathcal{O}\left( a^2 \right);
\eeq
where:
\beq
\label{eq_FF_soft_one}
\Sigma_S^{(1)}(y;\rho) \, = \, 
- \, \left[
\left( 1 \, + \, 2 \, \rho \right) \, 
\log\left( \frac{\rho}{1 \, + \, \rho} \right) \, + \, 2 
\right] \, \log(y).
\eeq
The following remarks are in order:
\begin{enumerate}
\item
In the massless limit $\rho \to 0^+$, the form factor contains
the singular term
\beq
- \, a \, \log(\rho) \, \log(y),
\eeq
with $\log(\rho)$ a mass singularity of collinear origin;
\item
In the no-recoil limit $\rho \to + \infty$
(equivalent to the limit $r \to 1^-$),
the form factor exactly vanishes,
\beq
\lim_{\rho \to + \infty} \Sigma_S^{(1)}(y;\rho) \, = \, 0,
\eeq
as could be expected on physical ground.
\end{enumerate}


\subsection{Soft Coefficient Function}

Let's now follow, in the present massive case, 
the standard factorization procedure
described in sec.$\,$\ref{sec_massless} for the 
simpler massless case.
According to eq.(\ref{eq_factor_expand}),
at first order in $a$ (equivalently, in $\alpha_S$), 
the sum of the first-order Coefficient function
$C_S^{(1)}(\rho)$
and Remainder function 
$\mathrm{Rem}_S^{(1)}(y;\rho)$,
is obtained by subtracting, 
from the first-order rate $E^{(1)}$,
the first-order form factor $\Sigma_S^{(1)}$:
\bea
\label{eq_CFplusRem}
&& C_S^{(1)}(\rho) \, + \, \mathrm{Rem}_S^{(1)}(y;\, \rho) 
\, = \,
E^{(1)}(y; \, \rho)  \, - \, \Sigma_S^{(1)}(y; \, \rho) \, =
\\
&=& \big( 1 \, + \, 2 \, \rho \big)
\bigg[ 
\, \frac{1}{2} \, \log^2\left( y \, + \, \rho \right) 
\, - \,
\, \frac{1}{2} \, \log^2\left( 1 \, + \, \rho \right)
\, - \,
\log\left( 1 + \frac{y}{\rho} \right) \log(y) \, +
\nonumber\\
&&\qquad\qquad\qquad\qquad\qquad\qquad\qquad\qquad
\, + \, \mathrm{Li}_2\left( \frac{\rho}{y \, + \, \rho} \right)
\, - \, \mathrm{Li}_2\left( \frac{\rho}{1 \, + \, \rho} \right)
\bigg] \, +
\nonumber\\
&+& \frac{1}{4} \, 
\log\left( \frac{y \, + \, \rho}{1 \, + \, \rho} \right)
+ \left[
\rho \left( 
2 \, \rho^2 \, + \, 3 \, \rho \, + \, \frac{1}{2}
\right) 
\,  - \, y \left( 1 \, - \, \frac{y}{4} \right) 
\right] 
\log\left( \frac{y+\rho}{1+\rho} \right) \, +
\nonumber\\
&+& \, \frac{1 \, - \, y}{y \, + \, \rho}
\left\{
\rho \left[
2 \, \rho^2 \, + \, \rho \, (y \, + \, 2)
\, - \, \frac{1}{6}  \left(2 \, y^2 \, - \, 7 \, y \, + \, 14 \right)
\right]
\, + \, \frac{y}{12} \left(2 \, y^2 \, - \, y \, - \, 31 \right)
\right\}.
\nonumber
\eea
Note that the last member of the above expression,
unlike the physical spectrum, does not diverge for $y \to 0^+$,
because of the subtraction of all the soft logs 
contained in the soft form factor;
the most singular terms for $y \to 0^+$ are
of the form $y \log(y).$

\begin{figure}[ht]
\begin{center}
\includegraphics[width=0.5\textwidth]{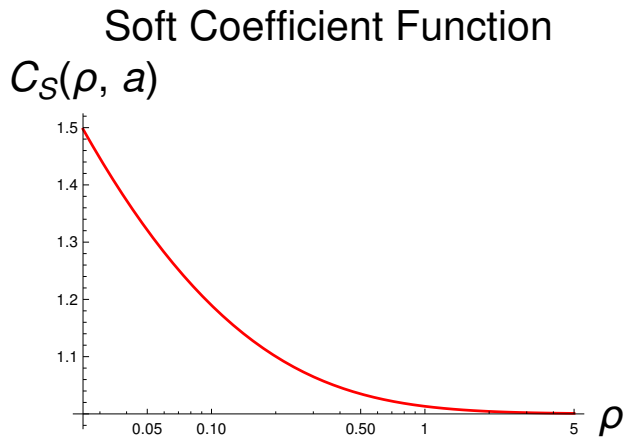}
\footnotesize
\caption{
\label{fig_PlotCS}
\it
Plot of Soft Coefficient Function 
$C_S(\rho;a)$ in first-order approximation
(see eq.(\ref{eq_C_S_complete})),
as a function of $\rho$, in the wide interval
$0.025 < \rho < 5$. 
The (horizontal) $\rho$-scale is logarithmic.
We have defined $a \equiv C_F \alpha_S(m_b) / \pi \cong 0.089$
for $\alpha_S(m_b)=0.21$.
A quite strong dependence of $C_S(\rho;a)$ for small $\rho$
is observed, as expected;
at $\rho=0.025$, the first-order correction is already about
50\% of the tree level value (which is one).
}
\end{center}
\end{figure}

The Soft coefficient function 
has the usual perturbative expansion
beginning with one:
\beq
C_S(\rho;a) \, = \, 1 \, + \, a \, C_S^{(1)}(\rho) \, + \, 
\mathcal{O}\left( a^2 \right).
\eeq
The first-order Soft Coefficient function is obtained by taking the limit $y\to 0^+$ of all members of eq.(\ref{eq_CFplusRem})
and taking into account that the Soft Remainder function
vanishes in this limit:
\bea
\label{eq_give_CS1}
C_S^{(1)}(\rho) &=& 
\lim_{y \to 0^+}
\left[
E^{(1)}(y;\rho)  \, - \, \Sigma_S^{(1)}(y;\rho) \,
\right] \, =
\\
&=& 
\big( 1 \, + \, 2 \rho \big) 
\left\{
\frac{1}{2} \,
\Big[
\log^2(\rho) \, - \, \log^2(1 \, + \, \rho)
\Big]
\, - \, \mathrm{Li}_2\left( \frac{\rho}{1 \, + \, \rho} \right)
\, + \, \frac{\pi^2}{6}
\right\} \, +
\nonumber\\ 
&+&
\left(
2 \, \rho^3 \, + \, 3 \, \rho^2 \, + \, \frac{\rho}{2} 
\, + \, \frac{1}{4}  
\right)
\Big[
\log(\rho) \, - \, \log(1 \, + \, \rho)
\Big] 
\, + \, 2 \, \rho(1 \, + \, \rho) \, - \, \frac{7}{3} .
\nonumber
\eea
Note that the first-order Soft Coefficient function 
$C_S^{(1)}(\rho)$ 
contains the double-logarithmic term of the strange mass
\beq
\frac{1}{2} \, \log^2(\rho),
\eeq
as well as the single-logarithmic term 
\beq
\frac{1}{4} \, \log(\rho),
\eeq
both diverging in the massless limit for the final quark, 
$\rho \to 0^+$.
The complete Soft Coefficient function up to first order,
\beq
\label{eq_C_S_complete}
C_S(\rho,a) \, = \, 1 \, + \, a \, C_S^{(1)}(\rho), 
\eeq
is plotted in fig.$\,$\ref{fig_PlotCS}.
%
%
%
%


\subsection{Soft Remainder Function}

\begin{figure}[ht]
\begin{center}
\includegraphics[width=0.5\textwidth]{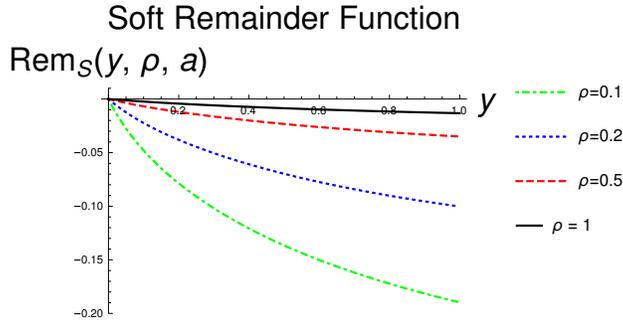}
\footnotesize
\caption{
\label{fig_PlotRemS}
\it
Plot of Soft Remainder Function 
$Rem_S(y;\rho;a)$ in first-order approximation,
eq.(\ref{eq_Soft_Rem_Funct}),
as a function $y$ in all its kinematic range,
$0  \le y \le 1$,
for four different values of $\rho$.
As shown in the figure:
$\rho=1$: black continuous line;
$\rho=0.5$: red dashed line; 
$\rho=0.2$: blue dotted line;
$\rho=0.1$: green dot-dashed line.
By reducing $\rho$ towards zero,
the Remainder Function becomes progressively
bigger in size.
}
\end{center}
\end{figure}

The Soft Remainder Function has the usual perturbative expansion beginning at first order:
\beq
\mathrm{Rem}_S(y;\rho;a) \, = \, a \, \mathrm{Rem}_S^{(1)}(y;\rho) 
\, + \, \mathcal{O}\left( a^2 \right).
\eeq
According to eq.(\ref{eq_CFplusRem}),
the Soft Remainder Function 
has the first-order (leading) term given by:
\bea
\label{eq_Soft_Rem_Funct}
\mathrm{Rem}_S^{(1)}(y; \rho) 
&=&
E^{(1)}(y;\rho)  \, - \, \Sigma_S^{(1)}(y;\rho) \, - \, C_S^{(1)}(\rho) \, =
\nonumber\\
&=&
\big( 1 \, + \, 2 \, \rho \big) 
\bigg\{
\frac{1}{2}  
\Big[ 
\log^2(y \, + \, \rho) \, - \, \log^2(\rho) 
\Big]
\, + \, \text{Li}_2\left( \frac{\rho }{y \, + \, \rho} \right)
\, - \, \frac{\pi^2}{6} \, +
\nonumber\\
&& \qquad\qquad\qquad\qquad\qquad\qquad\qquad
- \, \log(y) \Big[ \log(y \, + \, \rho)\, - \, \log(\rho) \Big]
\bigg\} \, +
\nonumber\\
&+&
\left( 2 \, \rho^3 \, + \, 3 \, \rho^2 \, + \, \frac{\rho}{2} \, + \, \frac{1}{4} \right) 
\Big[ \log (y \, + \, \rho) \, - \, \log (\rho) \Big] +
\nonumber\\
&-& \frac{y}{y \, + \, \rho}
\Bigg\{
\rho\left[
2 \, \rho^2 \, + \, \rho \, (y \, + \, 3) 
\, - \, \frac{1}{6} \left( 2 \, y^2 \, - \, 9 \, y \, + \, 9 \right) 
\right]
\, +
\nonumber\\
&& \qquad\qquad\qquad\qquad\qquad\qquad\qquad\qquad
+ \, \frac{y}{12} 
\left(
2 \, y^2 \, - \, 3 \, y \, - \, 30
\right)
\Bigg\} \, +
\nonumber\\   
&+& y \left( 1 \, - \, \frac{y}{4} \right) 
\Big[ 
\log(1 \, + \, \rho) \, - \, \log (y \, + \, \rho)
\Big]
\, - \, \frac{y}{4 \, (y \, + \, \rho)}.
\eea
By taking into account that: 
\beq
\text{Li}_2\left(1\right) \, = \, 
\sum_{n=1}^\infty \frac{1}{n^2} \, = \, \frac{\pi^2}{6},
\eeq
it is immediate to check the vanishing of the 
Soft Remainder function in the Born kinematics,
i.e.\ in the limit $y \to 0^+$ 
(in taking this limit, $\rho$ is kept constant and not zero: $\rho=\rho_0>0$).
The Soft Remainder Function is plotted in 
fig.$\,$\ref{fig_PlotRemS}
for different values of $\rho$.
%
%
%

Since
\beq
\mathrm{Rem}_S^{(1)}(y=1;\,\rho) \, = \, - \, C_S^{(1)}(\rho), 
\eeq
by reducing $\rho$ towards zero (from above, let's say, from $\rho=1$),
the corresponding Soft Remainder functions, which are all
strictly monotonically decreasing functions, 
and therefore all negative, become
progressively bigger in size.

We can conclude that the Soft Factorization scheme which we have constructed in this section, is very simple, but it
only works in the region
it is aimed at, namely $y \ll 1$ and $\rho = \mathcal{O}(1)$:
there is no bonus.
To consistently describe the small $y$, small-mass region $\rho \ll 1$,
we have to construct a more general factorization
scheme, based on a $QCD$ form factor which also resumes
small-$\rho$ effects, i.e.\ the collinearly enhanced terms.


\section{General Factorization}
\label{sec_gener_factor}

In this section we construct a general factorization
scheme, which correctly describes the soft region
\beq
0 \, < \, y \, \ll \, \rho \, = \, \mathcal{O}(1)
\qquad (\mathrm{soft\,\,region})
\eeq
and the (effectively) massless region 
\beq
0 \, < \, \rho \, \ll \, y \, \ll \, 1
\qquad (\mathrm{effectively\,\,massless\,\,region}),
\eeq
as well as the "transition region" 
\beq
0 \, < \, y \, \approx \, \rho \, \ll \, 1
\qquad (\mathrm{transition\,\,region}).
\eeq
As in the previous massless or soft factorization schemes,
the factorized event fraction is written as:
\beq
\label{eq_gener_factor}
E\left(y;\rho;\alpha_S\right)
\, = \, C\left(\rho;\alpha_S\right) \, 
\Sigma\left(y;\rho;\alpha_S\right) 
\, + \, 
\mathrm{Rem}\left(y;\rho;\alpha_S\right).
\eeq
In order to determine the Coefficient function,
$C(\rho;\alpha_S)$, as well as the Remainder function, 
$\mathrm{Rem}(y;\rho;\alpha_S)$,
to $\mathcal{O}\left(\alpha_S\right)$, 
we need to know the general $QCD$ form factor 
$\Sigma\left(y;\rho;\alpha_S\right)$,
also at first order in the coupling.
The latter cannot be evaluated directly, 
but it can be obtained by expanding
in powers of $\alpha_S$ the general resummed
form factor, as described in the next section.


\subsection{General $QCD$ Form Factor}

The general $QCD$ form factor,
resumming to all orders in $\alpha_S$
the infrared (soft and/or collinear) 
logarithms occurring in the perturbative expansion
of the event fraction,
has the following expression \cite{Aglietti:2007bp}:
\bea
\label{QCD_FF_gen}
&& \sigma_N\left(\rho,\alpha_S\right) \, = \,
\exp
\int\limits_0^1 dy \,
\Big[
(1-y)^{N-1} \, - \, 1
\Big]
\Bigg\{ \frac{1}{y}
\int\limits_{Q^2 y^2/(1+\rho)}^{Q^2 y^2/(y+\rho)}
\frac{dk_\perp^2}{k_\perp^2} \,
A\left[ \rho ; \alpha_S\left(k_\perp^2\right) \right] \, +  
\\
&&
+ \, B\left[ \alpha_S \left( \frac{Q^2y^2}{y + \rho} \right) \right]
\frac{1}{y + \rho}
\, + \, D\left[ \alpha_S \left( \frac{Q^2y^2}{1 + \rho} \right) \right]
\frac{1}{y}
\, + \, \Delta\left[ \alpha_S\left( \frac{Q^2y^2}{1 + \rho} \right) \right] 
\left(
\frac{1}{y} \, - \, \frac{1}{y + \rho}
\right)
\Bigg\}.
\nonumber
\eea
%
%
Note that, in the above equation, the term
proportional to $\Delta(\alpha_S)$ describes
soft-gluon emission off the strange quark line 
for $y \lsim \rho$.
As a consequence, this term identically vanishes
in the massless limit $\rho \to 0^+$.
We may say that the $B$ and $\Delta$ terms
are somehow "complementary", in the sense that one acts
in the kinematic region where the other does not.

Let us remark that the general form factor in eq.(\ref{QCD_FF_gen})
reduces to:
\begin{enumerate}
\item
the Heavy-to-Light form factor, 
eq.(\ref{QCD_FF_standard}),
in the massless limit $\rho \to 0^+$;
\item
the Soft form factor, eq.(\ref{QCD_FF_soft}), 
in the soft limit $y \to 0^+$, $\rho=$const$\ne 0$
or, more simply, $y/\rho \to 0^+$.  
\end{enumerate}	
%


\subsubsection{Form-factor expansion}

To compute the partially-integrated form factor 
$\Sigma = \Sigma(y; \rho; \alpha_S)$, expanded up to first
order in $\alpha_S$,
the only non-trivial integration involved is that one
of the term 
proportional to $a \, A^{(1)}(\rho)$, namely
\beq
\label{eq_non_trivial_integr}
\int\limits_y^1 \frac{d y'}{y'} 
\, \log\left( \frac{y' \, + \, \rho}{1 \, + \, \rho} \right).
\eeq
In the above expression,
the integration of the soft-gluon transverse
momentum squared $k_\perp^2$ has already been made.
In order to isolate the large infrared logarithms,
the expansion of the resummed form factor is conveniently written
--- out of the many possible forms --- as: 
\bea
\label{eq_Gen_FF_expanded}
\Sigma(y;\rho;a) &=& 1 \, + \, a \, A^{(1)}(\rho)
\left[
- \, \log\left( \frac{y+\rho}{1+\rho} \right) \log(y) 
\, + \, \frac{1}{2} \, \log^2\left( \frac{y+\rho}{1+\rho} \right)
\right. +
\nonumber\\
&& \qquad\quad
\left.
\, + \, \log(1 \, + \,\rho) \, \log\left( \frac{y+\rho}{1+\rho} \right)
\, + \, \mathrm{Li}_2\left( \frac{\rho}{y+\rho} \right) 
\, - \, \mathrm{Li}_2\left( \frac{\rho}{1+\rho} \right) 
\right] +
\nonumber\\
&+&
a \, D^{(1)} \, \log(y) 
\, + \, a \, \Delta^{(1)} \, 
\left[ 
\log(y) \, - \, \log\left( \frac{y+\rho}{1+\rho} \right)
\right]
\, + \, a \, B^{(1)} \, \log\left( \frac{y+\rho}{1+\rho} \right)
 \, +
\nonumber\\
&& 
\qquad\qquad\qquad\qquad\qquad\qquad\qquad
\,\, + \,\, \mathcal{O}\left(a^2\right).
\eea
The following remarks about the above equation
are in order:
\begin{enumerate}
\item
Unlike the massless form factor, 
eq.(\ref{QCD_FF_massless_ord1}) or (\ref{eq_QCD_FF_massless_first_ord}),
a $\log^2(y)$ term is absent in eq.(\ref{eq_Gen_FF_expanded}), because
of the regulating effect on collinear emissions of a non-vanishing
strange mass, i.e. $\rho \ne 0$, as discussed
in the Introduction.
Apart from this term, actually all the possible
quadratic and linear terms containing
$\log(y)$ and $\log(y+\rho)/(1+\rho)$
do appear in eq.(\ref{eq_Gen_FF_expanded}); 
\item
The arguments of the dilogarithms
are always smaller than, or equal to, one,
so these term are uniformly bounded by 
$\mathrm{Li}_2(1) = \pi^2/6 \simeq 1.64493$
--- namely a constant of order one.
\end{enumerate}
%
Checking that the first square bracket on the r.h.s.
of eq.(\ref{eq_Gen_FF_expanded})
is equal to the integral in (\ref{eq_non_trivial_integr}) 
is quite standard:
\begin{enumerate}
\item
One takes the derivatives with respect to $y$
of both expressions and checks that they
are equal;
\item
One checks that both expressions are equal 
for a particular value of $y$, such as for example 
the point $y=1$,
where the integral in (\ref{eq_non_trivial_integr}) vanishes.
\end{enumerate}
By replacing the explicit values of
the first-order coefficients,
one obtains for the first-order form factor,
\beq
\Sigma(y;\rho;a) 
\, = \, 1  \, + \, a \, \Sigma^{(1)}(y;\rho)
\, + \, \mathcal{O}\left(a^2\right),
\eeq
the explicit expression:
\bea
\label{eq_gen_FF_expand_1}
\Sigma^{(1)}(y;\rho) &=&  
\big( 1 \, + \, 2 \, \rho \big)
\Bigg[
- \, \log\left( \frac{y+\rho}{1+\rho} \right) \, \log(y) 
\, + \, \frac{1}{2} \, \log^2\left( \frac{y+\rho}{1+\rho} \right)
 \, +
\nonumber\\
&& \qquad\qquad\quad
\, + \, \log(1 \, + \,\rho) \, \log\left( \frac{y+\rho}{1+\rho} \right)
+ \, \mathrm{Li}_2\left( \frac{\rho}{y \, + \, \rho} \right)
- \, \mathrm{Li}_2\left( \frac{\rho}{1 \, + \, \rho} \right)
\Bigg] \, + 
\nonumber\\
&& \, - \, 2 \log(y) 
\, + \, \frac{1}{4} \log\left( \frac{y+\rho}{1+\rho} \right) .
\eea
The following remarks about the form factor above, 
eq.(\ref{eq_gen_FF_expand_1}), are in order:
\begin{enumerate}
\item
It reduces to the massless form factor, 
eq.(\ref{QCD_FF_massless_ord1}),
in the massless limit $\rho \to 0^+$,
and to the soft form factor, eq.(\ref{eq_FF_soft_one}), 
for $y/\rho \to 0^+ $, $\rho \gsim 1$; 
\item
It  vanishes in the limit of vanishing photon energy:
\beq
\lim_{y \to 1^-} \Sigma^{(1)}(y;\rho) \, = \, 0,
\eeq
where one has simply to take into account that 
\beq
\lim_{y \to 1^-} z(y;\rho) \, = \, 1. 
\eeq
\end{enumerate}


\subsection{General Coefficient function}

In order to evaluate the Coefficient 
and Remainder functions, 
the first step is to subtract,
from the first-order spectrum, 
eq.(\ref{comp_Ali_and_Greub}), the 
first-order $QCD$ form factor, eq.(\ref{eq_gen_FF_expand_1}).
That way, one obtains:
\bea
\label{CF_RemFun_Gen}
C^{(1)}(\rho) \, + \, \mathrm{Rem}^{(1)}(y;\rho) &=&
E^{(1)}(y;\rho) \, - \, \Sigma^{(1)}(y;\rho) \, =
\\
&=& \bigg[
\rho \left(
2 \, \rho^2 \, + \, 3 \, \rho \, + \, \frac{1}{2}
\right) 
\, - \, y \left( 1 \, - \, \frac{y}{4} \right)
\bigg] 
\log \left( \frac{y \, + \, \rho}{1 \, + \, \rho} \right) 
 \, +
\nonumber\\
&& + \, \frac{1 \, - \, y}{y \, + \, \rho} 
\bigg[
2 \, \rho^3 \, + \, \rho^2 (y \, + \, 2) 
\, - \, \frac{\rho}{6} 
\left( 2 \, y^2 \, - \, 7 \, y \, + \, 14 \right) \, +
\nonumber\\
&& \qquad\qquad\qquad\qquad\qquad\qquad\quad
\, + \, \frac{y}{12} 
\left( 2 \, y^2 \, - \, y \, - \, 31 \right)
\bigg].
\nonumber
\eea
The following remarks are in order:
\begin{enumerate}
\item
All soft logarithms --- namely all $\log(y)$ terms ---
exactly canceled by subtracting from the spectrum
the form factor, as it should, and as already happened 
in the (simpler) soft factorization scheme.
Actually, in the present case, unlike the soft one, 
the complete cancellation of the first three rows
on the r.h.s.\ of eq.(\ref{comp_Ali_and_Greub}) occurred
--- a large number of terms canceled;
\item
On the last member of eq.(\ref{CF_RemFun_Gen}),
the coefficient of the collinear logarithm 
--- namely the term $\log\left[(y + \rho)/(1+\rho)\right]$ --- 
is suppressed by
positive powers of $\rho$ or $y$, again as it should.
Note that this did not happen in the soft factorization scheme;
\item
It is immediately checked that the event fraction
minus the form factor exactly vanishes at 
the upper endpoint $y=1$.
\end{enumerate}
The first-order Coefficient function is given by:
\beq
\label{eq_Coef_Fun_First_Ord}
C^{(1)}(\rho) \, = \, 
\lim_{y \to 0^+} 
\Big[
E^{(1)}(y;\rho) \, - \, \Sigma^{(1)}(y;\rho)
\Big].
\eeq
By taking the above limit, one easily finds:
\beq
\label{eq_C1_rho}
C^{(1)}(\rho) \, = \, - \, \rho  
\left(
2 \, \rho^2 \, + \, 3 \, \rho \, + \, \frac{1}{2}
\right)
\Big[ 
\log(1 \, + \, \rho) \, - \, \log (\rho)
\Big]
\, + \, 2 \, \rho \, (1 \, + \, \rho) \, - \, \frac{7}{3}.
\eeq
Unlike the Soft Coefficient function, eq.(\ref{eq_give_CS1}), 
which, as we have shown, diverges like $\log^2(\rho)$ for $\rho \to 0^+$,
the general Coefficient function above has a finite value,
of order one, in the massless limit:
\beq
\label{eq_lim_C1_rho_to_zero0}
\lim_{\rho \to 0^+} C^{(1)}(\rho) \, = \, - \, \frac{7}{3}.
\eeq
Note that the most singular terms for $\rho \to 0^+$
in $C^{(1)}(\rho)$
are of the form $\rho \log(\rho)$
(just like the $\mathcal{O}(\alpha_S)$ correction
factor $K(\rho)$ to the inclusive width, see eq.(\ref{eq_Kfactor_btosdecay})).

\begin{figure}[ht]
\begin{center}
\includegraphics[width=0.5\textwidth]{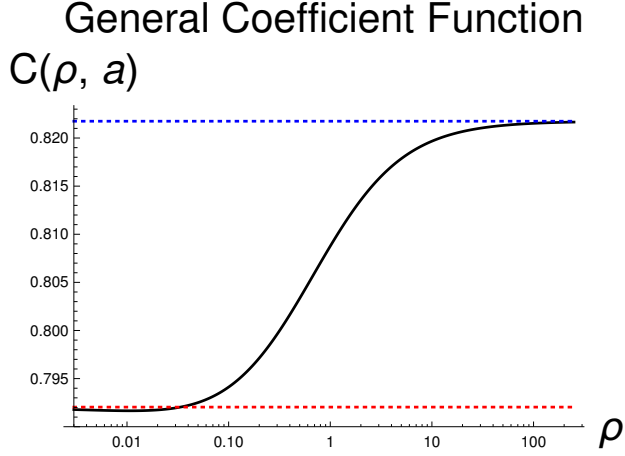}
\footnotesize
\caption{
\label{fig_PlotCF}
\it
Plot of the General Coefficient function $C(\rho;a)$ at first order,
eq.(\ref{eq_Coef_Fun_Massive}),
as a function of the mass parameter $\rho$, on a logarithmic scale.
We have defined $a \equiv C_F\alpha_S(m_b)/\pi \simeq 0.089$
for $\alpha_S(m_b)=0.21$.
The dotted red and blue lines represent
the asymptotic values of $C(\rho,a)$
for $\rho \to 0^+$ and $\rho \to + \infty$
respectively.
A rather mild dependence of $C(\rho;a)$ on $\rho$
is observed.
}
\end{center}
\end{figure}

The no-recoil limit ($\rho \to +\infty$) of the general Coefficient function
does not vanish, but is finite: 
\beq
\lim_{\rho \to +\infty} C^{(1)}(\rho) \, = \, - \, 2.
\eeq
Therefore, by increasing $\rho$ from zero to infinity,
the first-order coefficient function increases by $1/3$.
A plot of the complete Coefficient function
at first order,
\beq
\label{eq_Coef_Fun_Massive}
C(\rho;a) \, = \, 1 \, + \, a \, C^{(1)}(\rho),
\eeq
as a function of the mass parameter $\rho$, is given in 
fig.$\,$\ref{fig_PlotCF}.
We observe that $C(\rho;a)$ is basically a monotonically-increasing function of 
$\rho$,
with a rather mild dependence on this variable.
By comparing fig.$\,$\ref{fig_PlotCF} with fig.$\,$\ref{fig_PlotCS},
we notice a substantial stabilization 
of the Coefficient function
as far as the dependence
on $\rho$ is concerned; That is, generically speaking,
a "good new".
Furthermore, while the soft coefficient function
is always greater than one, the general coefficient function
is always smaller than one.

%
%
%
%


\subsection{General Remainder function}

The first-order Remainder function collects, by definition, 
all the $\mathcal{O}(\alpha_S)$ terms which are not
included neither in the Form Factor nor
in the Coefficient function:
\beq
\mathrm{Rem}^{(1)}(y;\rho) 
\, \equiv \, 
E^{(1)}(y;\rho) \, - \, \Sigma^{(1)}(y;\rho) \, - \, C^{(1)}(\rho).
\eeq
By construction (see eq.(\ref{eq_Coef_Fun_First_Ord})), 
it vanishes on Born kinematics:
\beq
\label{eq_Rem1_limit}
\lim_{y \to 0^+} \mathrm{Rem}^{(1)}(y;\rho) \, = \, 0
\qquad (\rho=\mathrm{const} > 0).
\eeq
The first-order Remainder function 
(the lowest non-vanishing order) explicitly reads:
\bea
\label{eq_Rem1_massive}
\mathrm{Rem}^{(1)}(y;\rho)&=& 
\rho \left(2 \, \rho^2 \, + \, 3 \, \rho \, + \, \frac{1}{2} \right)  
\log\left( 1 \, + \, \frac{y}{\rho} \right) 
\, - \, y \left( 1 \, - \, \frac{y}{4} \right)
\log\left( \frac{y \, + \, \rho}{1 \, + \, \rho} \right) \, +
\nonumber\\
&-& 2 \, \rho^2 \, y \, - \, \rho \, y \, (3 \, - \, y)
\, - \, \frac{y}{6} \left( 4 \, y^2 \, - \, 9 \, y \, - \, 9 \right) \, +
\nonumber\\
&+&\frac{ y^2 }{4 \, (y \, + \, \rho)}
\left( 2 \, y^2 \, - \, 5 \, y \, + \, 4 \right)
\, - \, \frac{y}{4 \, ( y \, + \, \rho )}.
\eea
%
%
\begin{figure}[ht]
\begin{center}
\includegraphics[width=0.5\textwidth]{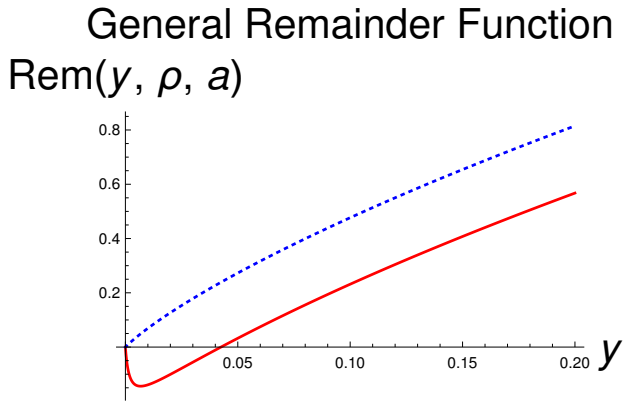}
\footnotesize
\caption{
\label{fig_PlotRem}
\it
The continuous red line is the plot of the General Remainder function 
$\mathrm{Rem}(y;\rho;a)$ in first order approximation,
eq.(\ref{eq_Rem1_massive}),
as a function of $y$, in the small-$y$ range 
$0 \le y \le 0.2$, for $\rho=0.002$.
For comparison (see text), we have also plotted the subtracted
Remainder function, eq.(\ref{eq_def_Rem1_sub}) 
--- the blue dotted line.
}
\end{center}
\end{figure}
%
%
It is immediate to check that the r.h.s.\ of the above equation
vanishes for $y \to 0^+$ ($\rho=\mathrm{const}>0$), 
as all terms are explicitly proportional to
$y$ or to higher powers of $y$, or are proportional to
$\log (1 + y/\rho)$, which is also $\mathcal{O}(y)$. 

Note that the second and the third row
on the r.h.s.\ of eq.(\ref{eq_Rem1_massive}) 
have been rewritten by means of a partial
fractioning with respect to $\rho$.
The general Remainder function is plotted in 
fig.$\,$\ref{fig_PlotRem}.
Similarly to the General Coefficient function,
also the General Remainder function
has a mild dependence on the mass parameter $\rho$.


\section{Improved Factorization Scheme}
\label{sec_improv_factor_scheme}

In the previous section we have constructed
a factorization scheme,
given by eqs.(\ref{eq_gener_factor}), (\ref{QCD_FF_gen}),
(\ref{eq_C1_rho}) and (\ref{eq_Rem1_massive}),
which correctly works in the threshold region $(y \ll 1)$ 
in the massive case $(\rho \ne 0)$
so long as,
in taking the limit $y\to 0^+$, the mass parameter $\rho$
is kept constant and not zero: $\rho = \rho_0 > 0$.
However, it is also natural to ask ourselves
what happens if we take the massless limit
$\rho \to 0^+$ in our massive factorization formula.
We have seen that both the massive Coefficient
and Remainder functions have a finite limit
for $\rho \to 0^+$, while the soft Coefficient
and Remainder functions diverge like $\log^2(\rho)$
in the same limit.
Therefore, as expected, a substantial improvement is obtained
by going from the soft factorization scheme
to the general one, again as far as the massless limit $\rho\to 0^+$
is concerned.
The problem then is:
by taking the massless limit
of our general factorization formula,
do we obtain the same Form Factor,
the same Coefficient and Remainder functions of
the massless factorization scheme,
i.e.\ of the standard factorization scheme applied to the massless
event fraction $E_0$, described in 
sec.$\,$\ref{sec_massless}, or we do not?

Let us begin our analysis by studying the simpler object
occurring in the factorization process,
namely the Coefficient function.
The massless limit in eq.(\ref{eq_lim_C1_rho_to_zero0}),
\beq
\label{eq_lim_C1_rho_to_zero}
\lim_{\rho \to 0^+} C^{(1)}(\rho) \, = \, - \, \frac{7}{3},
\eeq
does not coincide with the massless Coefficient function, 
i.e.\ evaluated in the usual factorization of the massless 
spectrum ($\rho=0$ from the very beginning: see sec.\ref{sec_massless}):
\beq
C_0^{(1)} \, = \, \, - \, \frac{31}{12}.
\eeq
To obtain the limiting Coefficient function in 
eq.(\ref{eq_lim_C1_rho_to_zero}), 
one has to add to $C_0^{(1)}$ the constant $1/4$:
\beq
\label{cfr_CF}
\lim_{\rho\to 0^+} C^{(1)}(\rho) 
\, = \,
C_0^{(1)} \, + \, \frac{1}{4}.
\eeq
We can say that the following two operations 
do not commute with each other:%
\footnote{
Non-commuting phenomena also appear
in the factorization of
fragmentation functions of heavy quarks
\cite{Gaggero:2022hmv}.}
 
\begin{enumerate}
\item
Taking the massless limit
$\rho \to 0^+$ of the event fraction;
\item
Factorizing the spectrum into a form factor,
a coefficient and a remainder functions.
\end{enumerate}
A similar problem, in particular, 
a "specular mismatch", also occurs with the 
General Remainder function.
In the massless limit $\rho \to 0^+$,
the general Remainder function becomes:
\beq
\lim_{\rho \to 0^+} \mathrm{Rem}^{(1)}(y;\rho)
\, = \,
\, - \, y \left( 1 \, - \, \frac{y}{4} \right) \log(y)
\, - \, \frac{y}{12} \left( 2 \, y^2 \, - \, 3 \, y \, - \, 30 \right)
\, - \, \frac{1}{4}
\qquad (y=\mathrm{const}>0). 
\eeq
The massless limit above does not vanish in the 
Born-kinematic limit $y \to 0^+$:
\beq
\lim_{y \to 0^+} 
\left[
\lim_{\rho \to 0^+} \mathrm{Rem}^{(1)}(y;\rho)
\right]
\, = \, - \, \frac{1}{4}.
\eeq
Actually, "symmetrically" with respect to the case of the 
General Coefficient function,
the limiting General Remainder function is equal
to the massless one minus the constant $1/4$:
\beq
\label{cfr_Rem}
\lim_{\rho \to 0^+} \mathrm{Rem}^{(1)}(y;\rho)
\, = \, 
\mathrm{Rem}_0^{(1)}(y) \, - \, \frac{1}{4}.
\eeq
By comparing eq.(\ref{cfr_CF}) with eq.(\ref{cfr_Rem}),
we find that, by taking the massless limit
of the general massive factorization formula, 
the constant $1/4$ is moved from the Remainder function
to the Coefficient function. 
As already noted, 
the conclusion is that it makes a difference to take the massless limit
$\rho \to 0^+$ {\it before} Factorization
or {\it after} Factorization of the spectrum.
The problem does not originate from the $QCD$ form factor
which, as we have shown, has a smooth behavior in the massless limit
$\rho \to 0^+$,
so it necessarily originates from the 
splitting of the non-logarithmic terms
between the Coefficient function and the Remainder function.

By looking at the plot of the 
Remainder function in fig.$\,$\ref{fig_PlotRem}
for $\rho \ll 1$,
one finds a small dip for very small $y$.
The latter is produced by the last term on the r.h.s.\ of 
eq.(\ref{eq_Rem1_massive}), namely the term
\beq
\label{non_comm_term}
- \, \frac{y}{4 \, (y \, + \, \rho)}.
\eeq
Indeed the dip completely disappears in the subtracted Remainder function,
\beq
\label{eq_def_Rem1_sub}
\overline{\mathrm{Re}}^{(1)}(y;\rho)
\, \equiv \,
\mathrm{Re}^{(1)}(y;\rho) \, + \, \frac{y}{4 \, (y \, + \, \rho)},
\eeq
in which the above term has been removed by hand
(see fig.$\,$\ref{fig_PlotRem}).
Therefore we can conclude that the finite mismatch
which we have found, 
originates from this term only,
on which we then focus our attention from now on.

Formally, for any strictly positive (and fixed) value of 
$\rho$, whatever small,
the ratio (\ref{non_comm_term})
vanishes in the threshold limit $y\to 0^+$, so this term
is naturally relegated into the massive Remainder function.
However for {\it very small} values of the mass parameter $\rho$, 
\beq
0 \, < \, \rho \, \ll \, 1,
\eeq
the term (\ref{non_comm_term}) is {\it small} 
only in a {\it tiny} 
region of the kinematic variable $y$, 
namely the region
\beq
\label{eq_very_small_y}
0 \, < \, y \, \ll \, \rho 
\qquad (\mathrm{small} \,\, y),
\eeq
where:
\beq
- \, \frac{y}{4 \, (y \, + \, \rho)} 
\, \approx \, - \, \frac{y}{4 \, \rho} 
\, \ll \, 1.
\eeq
Therefore only in the small $y$ region (\ref{eq_very_small_y}) (small compared to $\rho$),
the term in (\ref{non_comm_term}) is reasonably inserted into the Remainder function.
In the complementary, large $y$ region
(again in a strong inequality sense and again compared to $\rho$),
\beq
0 \, < \, \rho \, \ll \, y 
\qquad (\mathrm{large} \,\, y),
\eeq
this term is approximately equal to a constant of order one,
\beq
- \, \frac{y}{4 \, (y \, + \, \rho)} \, \approx \, - \, \frac{1}{4},
\eeq
so it would be reasonable to relegate it inside
the Coefficient function, and no more inside the Remainder function.

Mathematically, the problem originates from the
fact that:
\beq
\lim_{y \to 0^+} \, - \, \frac{y}{4 \, (y \, + \, \rho)}
\, = \, 0,
\eeq
while:
\beq
\lim_{\rho \to 0^+} \, - \, \frac{y}{4 \, (y \, + \, \rho)}
\, = \, - \, \frac{1}{4}.
\eeq
The ratio (\ref{non_comm_term}) is the only term in the Remainder function, eq.(\ref{eq_Rem1_massive}),
for which the limits $y\to 0^+$ and $\rho\to 0^+$
do not commute with each other, as:
\beq
\lim_{\rho \to 0^+} 
\left[
\lim_{y \to 0^+} \, - \, \frac{y}{4 \, (y \, + \, \rho)}
\right]
\, = \, \lim_{\rho\to 0^+} 0 \, = \, 0;
\eeq
while:
\beq
\lim_{y \to 0^+} 
\left[
\lim_{\rho \to 0^+} \, - \, \frac{y}{4 \, (y \, + \, \rho)}
\right]
\, = \, \lim_{y \to 0^+} \left( - \, \frac{1}{4} \right) 
\, = \, - \, \frac{1}{4} \, \ne \, 0.
\eeq
On the contrary, for the terms of the form $y^n/(y+\rho)$
with $n>1$, appearing on the r.h.s.\ of eq.(\ref{eq_Rem1_massive}), 
the two limits $y \to 0^+$ and $\rho \to 0^+$ do commute
with each other.
Note also that more complicated non-commuting terms,
such as for example $y^2/(y + \rho)^2$ or $y^2/(y^2 + \rho^2)$, do not appear on the r.h.s.\ of 
eq.(\ref{eq_Rem1_massive}).
A crude solution to this problem is 
to define:
\begin{enumerate}
\item
{\it An Improved $(I)$ Coefficient function,} 
by adding  to the massive Coefficient function,
$C^{(1)}(\rho)$ in eq.(\ref{eq_C1_rho}),
the term (\ref{non_comm_term}) 
when the latter is large,
i.e.\ for $y>\rho$:
\beq
\label{eq_improv}
C_I^{(1)}(y;\rho) 
\, \equiv \,
C^{(1)}(\rho) \, - \, \frac{y}{4 \, (y \, + \, \rho)} \, 
\theta(y \, - \, \rho);
\eeq
where $\theta(x) \equiv 1$ for $x>0$
and zero otherwise is the standard Heaviside
unit step function;
\item
{\it An Improved Remainder function,}
by adding to the subtracted Remainder function,
$\overline{\mathrm{Re}}^{(1)}(y;\rho)$ in
eq.(\ref{eq_def_Rem1_sub}),
the term (\ref{non_comm_term})
when the latter is small, i.e.\
in the complementary case $y < \rho$:
\beq
\mathrm{Re}_I^{(1)}(y;\rho) 
\, \equiv \, 
\overline{\mathrm{Re}}^{(1)}(y;\rho)
\, - \, \frac{y}{4 \, (y \, + \, \rho)} \, \theta(\rho \, - \, y) .
\eeq
\end{enumerate}
Note that:
\beq
C_I^{(1)}(\rho;y) \, + \, \mathrm{Re}_I^{(1)}(y;\rho) 
\, \equiv \, 
C^{(1)}(\rho) \, + \, \mathrm{Re}^{(1)}(y;\rho),
\eeq
i.e., with the present Improvement, we have simply made a rearrangement
of terms among the Coefficient function and the
Remainder function.
Since the $QCD$ form factor $\Sigma$ equals unity in the free limit,
\beq
\Sigma(y;\rho;\alpha_S) \, = \, 
1 \, + \, \mathcal{O}\left( \alpha_S \right),
\eeq
it follows that the Improved resummed formula,
i.e.\ eq.(\ref{eq_gener_factor}) with the Improved Coefficient and Remainder functions,
coincides with the standard resummed formula or the
fixed-order event fraction to $\mathcal{O}\left( \alpha_S \right)$.

An important point is that  our improvement 
{\it necessarily} introduces a dependence on $y$ in 
the massive Coefficient function. 
In order to have a Coefficient function with a
dependence on $y$ which is as simple as possible,
one can split the ratio (\ref{non_comm_term}) as:
\beq
- \, \frac{y}{4 \, (y \, + \, \rho)} \, = \, 
- \, \frac{1}{4} \, + \, \frac{\rho}{4 \,(y \, + \, \rho)}.
\eeq
When $y>\rho$, i.e.\ when the ratio on the l.h.s.\ above
is large, one inserts the constant $-1/4$ 
inside the Improved Coefficient
function, while the (small) fraction 
$\rho/ ( 4 (y \, + \, \rho) )$ 
is inserted in the Improved Remainder function.
Therefore one can also define the simpler Improved
Coefficient function 
\beq
\label{eq_new_improv}
C_I^{(1)}(y;\rho) 
\, \equiv \, 
C^{(1)}(\rho) \, - \, \frac{1}{4} \, \theta(y-\rho) ;
\eeq
together with the Improved Remainder function
\beq
\label{eq_new_improv_2}
\mathrm{Re}^{(1)}_I(y;\rho) 
\, \equiv \, 
\overline{\mathrm{Re}}^{(1)}(y;\rho)
\, - \, \frac{y}{4 \, (y \, + \, \rho)} \, \theta(\rho-y) 
\, + \, \frac{\rho}{4 \, (y \, + \, \rho)} \, \theta(y-\rho) .
\eeq

\noindent
Let us remark that, even though the non-commuting term
(\ref{non_comm_term}) is numerically rather small in size,
it has its own relevance on the theoretical side.
We also expect the presence 
of similar non-commuting terms to be
generic in heavy-to-heavy decays, i.e.\
not to be restricted to the rare $B \to X_s \gamma$
decays. Furthermore, in different processes the size
of non-commuting terms can be numerically larger.


\subsection{Smoothing the transition}

The Improved factorization scheme, which we have constructed
in the previous section,
can be further improved by eliminating the discontinuities 
in the Coefficient and Remainder functions
produced by the $\theta$-functions.
We can regularize the discontinuities by means of smooth functions 
with a similar step behavior to the $\theta$-functions, 
such as for example the sigmoids (see figs.\ref{fig_Plotphi2}
and \ref{fig_PlotPhi1}):
\beq
\label{eq_define_sigmoid}
\theta(x)
\quad
\mapsto
\quad
S_\Delta(x) \, \equiv \, \frac{1}{\sqrt{2\pi} \, \Delta}
\int\limits_{- \infty}^x 
\exp\left( - \, \frac{t^2}{2 \, \Delta^2} \right) \, dt; 
\eeq
where $\Delta>0$ is a parameter specifying the 
size of the $x$-interval, centered around $x=0$, 
where most of the variation
of the function $S_\Delta(x)$ occurs.
It holds indeed:
\beq
S_\Delta(x=+\Delta) \, - \, S_\Delta(x=-\Delta) \, \simeq \, 0.68.
\eeq
Furthermore, in a weak sense:
\beq
\lim_{\Delta \to 0^+} \sigma_\Delta(x) \, = \, \theta(x),
\eeq
so we recover the previous case by sending
the auxiliary parameter $\Delta$ to zero.
As well known from statistics, $\Delta$ is the
standard deviation $\sigma$ of the Gaussian
distribution function inside the integral
on the r.h.s.\ of eq.(\ref{eq_define_sigmoid}).

It is immediate to check the following
basic properties of the function $S_\Delta(x)$:
\beq
\lim_{x\to-\infty} S_\Delta(x) \, = \, 0;
\qquad
\lim_{x\to+\infty} S_\Delta(x) \, = \, 1;
\qquad
S_\Delta(0) \, = \, \frac{1}{2};
\qquad
S_\Delta(x) \, + \, S_\Delta(-x) \, \equiv \, 1.
\eeq
The smoothed Improved Coefficient and Remainder functions 
are simply obtained from the previous improved ones,
eqs.(\ref{eq_new_improv}) and (\ref{eq_new_improv_2}),
by replacing the $\theta$-functions with $S_\Delta$-functions
with the same arguments:
\bea
\label{eq_smooth_improv}
C_I^{(1)}(y;\rho) &\equiv& C^{(1)}(\rho) \, - \, 
\frac{1}{4} \, S_\Delta(y-\rho) ;
\nonumber\\
\mathrm{Re}_I^{(1)}(y;\rho) &\equiv& 
\overline{\mathrm{Re}}^{(1)}(y;\rho)
 \, + \, \frac{\rho}{4 \, (y+\rho)} \, S_\Delta(y-\rho) 
 \, - \, \frac{y}{4 \, (y+\rho)} \, S_\Delta(\rho-y) .
\eea
It is natural to assume $\Delta$ to be a function
of $\rho$:
\beq
\Delta \, = \, \Delta(\rho),
\eeq
together with $\Delta < \rho$.
In practice, for the numerical value of $\Delta$, one can take a fraction of $\rho$, such as for example:
\beq
\Delta \, = \, \frac{\rho}{3} 
\quad 
\mathrm{or} \quad \frac{\rho}{2}.
\eeq	


\subsection{Partition of Unity}

\begin{figure}[ht]
\begin{center}
\includegraphics[width=0.5\textwidth]{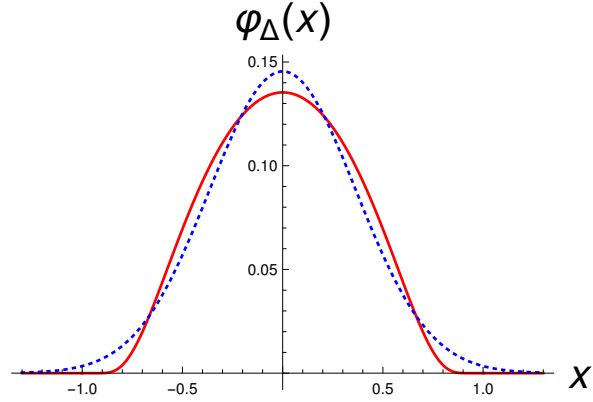}
\footnotesize
\caption{
\label{fig_Plotphi2}
\it
Plot of the function $\varphi_\Delta(x)$
for $\Delta=1$ in the range $-1.3 \le x \le 1.3$
(red continuous line).
As discussed in the main text, and as can also be seen
from the figure, this function is smooth
and identically equal to zero for $|x| \ge 1$.
For comparison, we have also plotted a Gaussian 
(blue dotted line) with the same normalization 
(integral over the real line)
and with standard deviation $\sigma=0.365$.
}
\end{center}
\end{figure}

One may wish to have an Improved
Coefficient function which is as
similar as possible to the standard one.
Actually, it is possible to construct an Improved 
Coefficient function which is: 
\begin{enumerate}
\item
Exactly equal to the massive coefficient function
$C^{(1)}(\rho)$, eq.(\ref{eq_C1_rho}), 
in the small-$y$ region
\beq
y \, < \, \rho \, - \, \Delta;
\eeq
\item
Independent of $y$ and with the correct
massless limit $\rho \to 0^+$, 
namely $C^{(1)}_0$ in eq.(\ref{eq_give_C1}), 
in the large-$y$ region
\beq
y \, > \, \rho \, + \, \Delta.
\eeq
\end{enumerate}
This problem can be solved
by means of the so-called Partition of Unity,
a general analytic method in real geometry \cite{bo}.
Let us consider the function
\beq
\varphi_\Delta(x) \, \equiv \, 
\exp\left( 
- \, \frac{1}{\Delta + x} \, - \, \frac{1}{\Delta - x} 
\right)
\qquad \mathrm{for} \,\,\, |x| \, < \, \Delta
\eeq
and zero otherwise.
In fig.$\,$\ref{fig_Plotphi2} we plot this function
for $\Delta=1$ (the red continuous line).
Note that the function $\varphi_1(x)$ is not qualitatively very different
from a Gaussian with standard deviation $\sigma=0.365 \approx \Delta/3$ (blue dotted line), 
even though the latter formally has support on the entire real line.
%
%
%
%

\begin{figure}[ht]
\begin{center}
\includegraphics[width=0.5\textwidth]{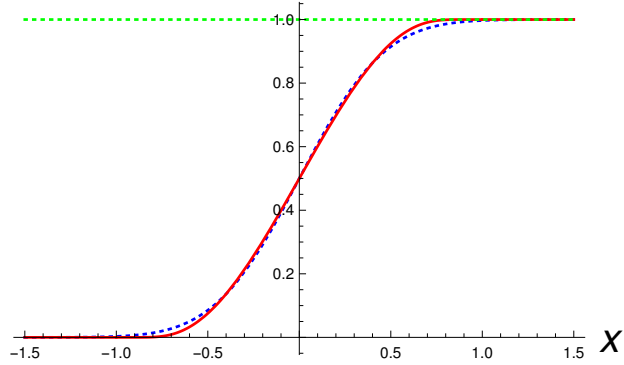}
\footnotesize
\caption{
\label{fig_PlotPhi1}
\it
The red continuous line is the plot of the function $\Phi_\Delta(x)$
for $\Delta=1$ in the range $-1.5 < x < 1.5$.
As discussed in the main text and as can also be seen
from the figure, this function is smooth, identically equal to zero
for $x \le -1$, identically equal to one
for $x \ge 1$, and strictly monotonically increasing
in the "transition region" $-1 \le x \le 1$.
The green dotted line represents the upper horizontal 
asymptote $y=1$.
The blue dotted line is the plot of the sigmoid, eq.(\ref{eq_define_sigmoid}),
with standard deviation $\Delta=0.365$.
Note that the two curves are barely distinguishable.
}
\end{center}
\end{figure}

It is immediate to check that $\varphi_\Delta(x)$ is an even function of $x$,
\beq
\varphi_\Delta(-x) 
\, = \, 
\varphi_\Delta(x).
\eeq
By explicitly computing the derivatives 
of $\varphi_\Delta(x)$ of all orders at $x=\pm\Delta$,
it is straightforward to check that this function
is infinitely smooth on the real line, 
\beq
\varphi_\Delta \, \in \, C^\infty(\RR).
\eeq
Given the normalization constant
\beq
K \, \equiv \, 
\int\limits_{-\Delta}^{+\Delta} \varphi_\Delta(x) \, dx \, > \, 0,
\eeq
let us define the function
\beq
\label{eq_def_Phi_Delta}
\Phi_\Delta(x) \, \equiv \, 
\frac{1}{K} \int\limits_{-\infty}^x \varphi_\Delta(x') \, dx'.
\eeq
Being the primitive of an infinitely smooth function,
also $\Phi_\Delta$ is a $C^\infty(\RR)$ function. 
This function also satisfies the following properties:
\begin{enumerate}
\item
It is identically equal to zero for small $x$, precisely:
\beq
\Phi_\Delta(x) \, \equiv \, 0 
\quad \mathrm{for} \,\, x \, < \, - \, \Delta;
\eeq
\item
It is identically equal to one for large $x$,
\beq
\Phi_\Delta(x) \, \equiv \, 1 
\quad \mathrm{for} \,\, x \, > \, + \, \Delta;
\eeq
\item
It is strictly monotonically increasing in the interval
\beq 
- \, \Delta \, < \, x \, < \, + \, \Delta.
\eeq
\end{enumerate}
The function $\Phi_\Delta(x)$ is plotted in fig.$\,$\ref{fig_PlotPhi1}
for $\Delta=1$.

\noindent
Note that the function 
\beq
\Psi_\Delta(x) \, \equiv \, \Phi_\Delta(x) \, - \, \frac{1}{2} 
\eeq
is odd in $x$,
\beq
\label{eq_Psi_odd}
\Psi_\Delta(-x) \, = \, - \, \Psi_\Delta(x).
\eeq
It vanishes indeed at $x=0$.

%
%
%
%

\begin{figure}[ht]
\begin{center}
\includegraphics[width=0.5\textwidth]{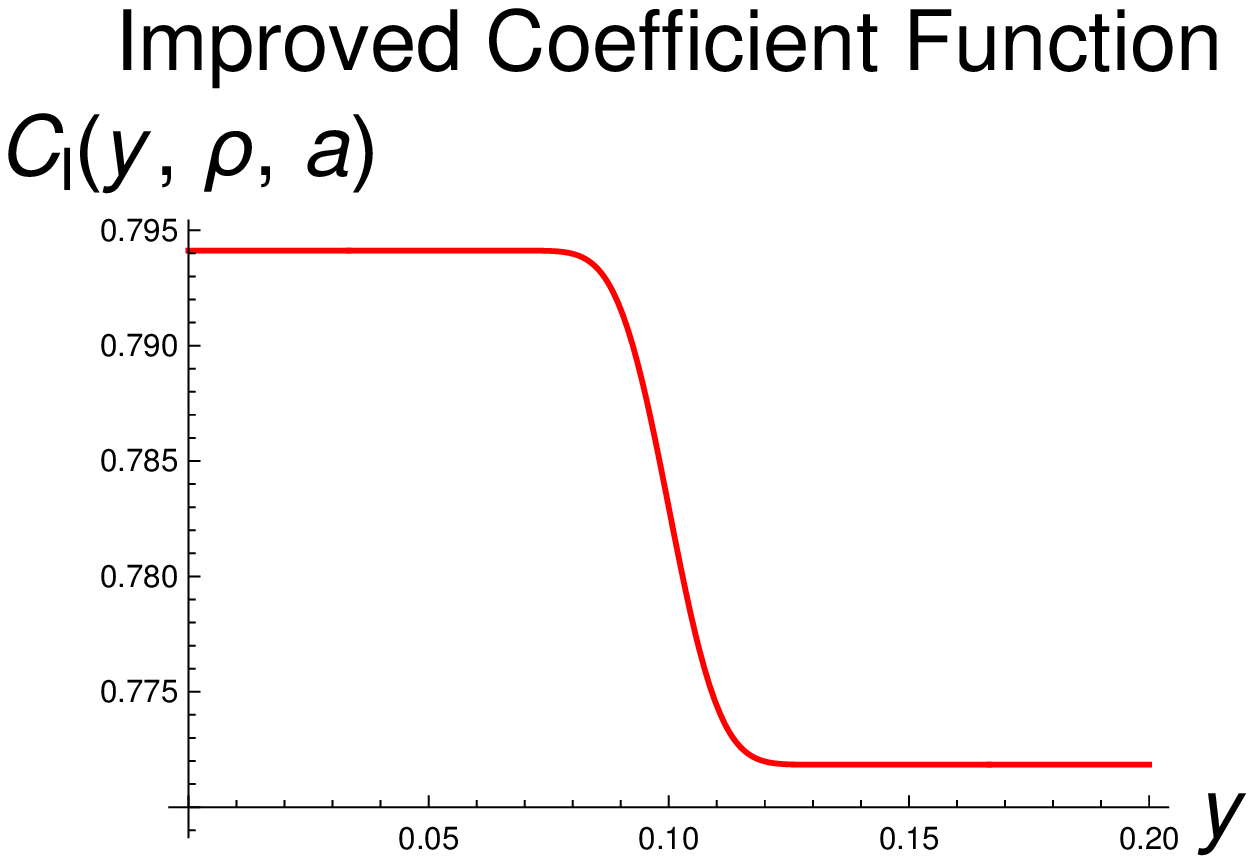}
\footnotesize
\caption{
\label{fig_PlotCimpr}
\it
Plot of the Improved Coefficient Function $C_I(y;\rho;a)$
at first order in $a$,
eq.(\ref{eq_Cimpr_best}),
for $\rho=0.1$ and $\Delta = 2/3 \, \rho \simeq 0.067$, 
as a function of $y$, in the small-$y$ region $0 \le y \le 0.2$.
The step behavior,
as well as the smooth transition around $y=0.1$
of total width $2\Delta$,
both induced by the partition of unity,
are clearly visible.
}
\end{center}
\end{figure}

\begin{figure}[ht]
\begin{center}
\includegraphics[width=0.5\textwidth]{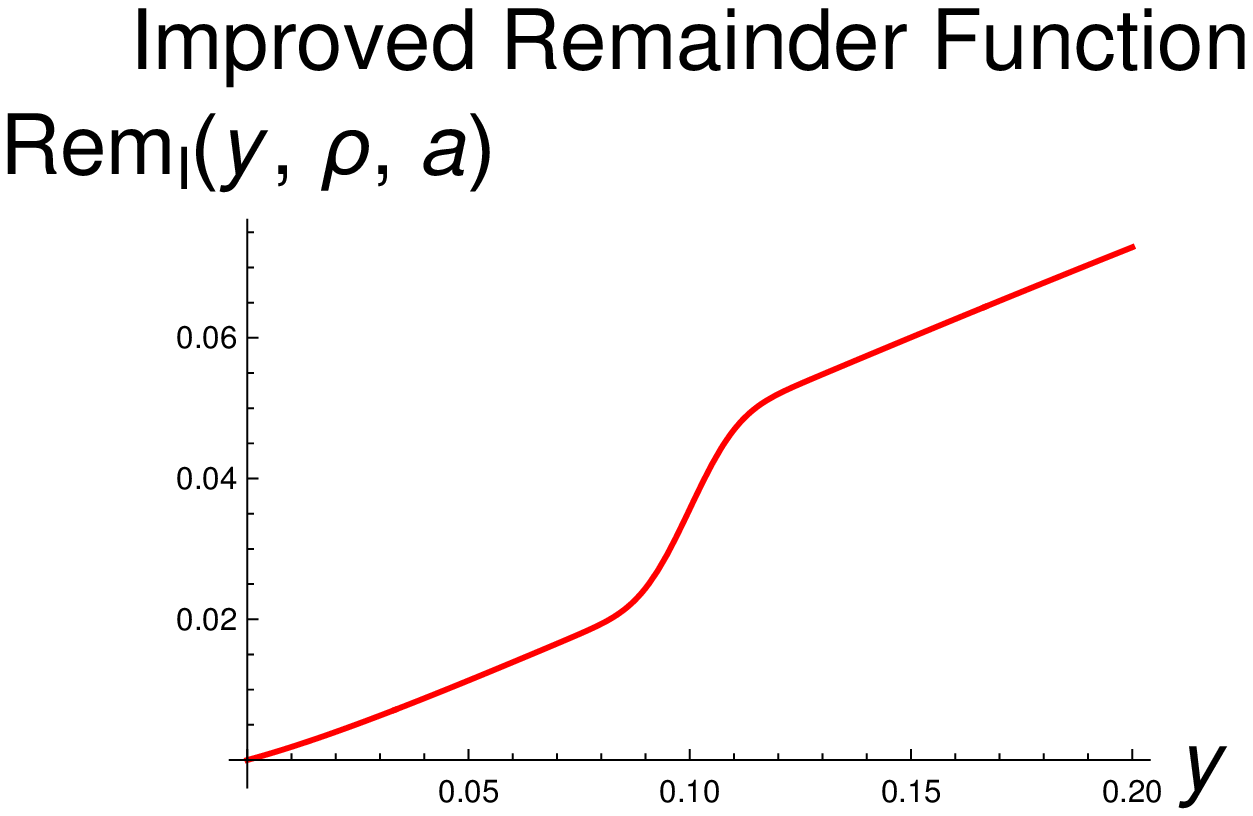}
\footnotesize
\caption{
\label{fig_PlotRemimpr}
\it
Plot of the Improved Remainder Function, 
$\mathrm{Rem}_I(y;\rho;a)$, at first order in $a$,
eq.(\ref{eq_Remimpr_best}),
as a function of $y$, in the small-$y$ region 
$0 \le y \le 0.2$,
for $\rho=0.1$ and $\Delta=2/3 \rho \simeq 0.067$.
The steeper rise around $y=0.1$,
induced by the partition of unity,
is clearly visible.
}
\end{center}
\end{figure}

\noindent
From eq.(\ref{eq_Psi_odd})
the fundamental relation follows:
%
%
\beq
\label{eq_fundam_relat}
\Phi_\Delta(x) \, + \, \Phi_\Delta(-x) \, \equiv \, 1,
\quad x \, \in \, \RR.
\eeq
The required Improved Coefficient and Remainder functions 
are simply obtained by replacing, 
inside eqs.(\ref{eq_smooth_improv}), 
the $S_\Delta$-functions with $\Phi_\Delta$-functions
with the same arguments:
\bea
\label{eq_Cimpr_Remimpr_best}
C_I^{(1)}(y;\rho) &\equiv& C^{(1)}(\rho) \, - \, 
\frac{1}{4} \, \Phi_\Delta(y \, - \, \rho) ;
\\
\mathrm{Re}^{(1)}_I(y;\rho) &\equiv& 
\overline{\mathrm{Re}}^{(1)}(y;\rho)
 \, + \, \frac{\rho}{4 \, (y \, + \, \rho)} \, 
 \Phi_\Delta(y \, - \, \rho)
 \, - \, \frac{y}{4 \, (y \, + \, \rho)} \, 
 \Phi_\Delta(\rho \, - \, y) .
\nonumber
\eea
The Improved Coefficient function at first order in $a$
(or in $\alpha_S$),
\beq
\label{eq_Cimpr_best}
C_I(y;\rho;a) \, = \, 1 \, + \, a \, C_I^{(1)}(y;\rho),
\eeq
is plotted in fig.$\,$\ref{fig_PlotCimpr}.
%
%
%
%
The Improved Remainder function at first order,
\beq
\label{eq_Remimpr_best}
\mathrm{Rem}_I(y;\rho;a) \, = \, a \, \mathrm{Rem}_I^{(1)}(y;\rho),
\eeq
is plotted in fig.$\,$\ref{fig_PlotRemimpr}.
%
%
%
%
Let's now prove that the above Improved Coefficient
function possesses all the properties we were looking for.
When $y<\rho-\Delta$, the function $\Phi_\Delta(y-\rho)$
identically vanishes, so that:
\beq
C_I^{(1)}(y;\rho) \, \equiv \, C^{(1)}(\rho)
\qquad (y \, < \, \rho \, - \, \Delta).
\eeq
When $y>\rho+\Delta$, the function $\Phi_\Delta(y-\rho)$
is identically equal to one, so that:
\beq
C_I^{(1)}(y;\rho) \, \equiv \, C^{(1)}(\rho) 
\, - \, \frac{1}{4}
\qquad
(y \, > \, \rho \, + \, \Delta). 
\eeq
According to eq.(\ref{cfr_CF}),
\beq
\lim_{\rho\to 0^+} C^{(1)}(\rho) \, - \, \frac{1}{4}
\, = \, C_0^{(1)},
\eeq
so that, in the massless limit $\rho\to 0^+$, we recover the
coefficient function $C_0^{(1)}$ of the
standard factorization of the massless spectrum
(i.e.\ the case where the limit $\rho\to 0^+$ is taken {\it before} the factorization
of the perturbative spectrum
into a form factor, a coefficient and a remainder functions). 


\subsection{Generalization}
\label{new_sec_gen}
Up to now we have explicitly considered only the
contribution of the (leading) operator $O_7$
to the radiative $B \to X_s \gamma$ decay,
contained in the effective non-leptonic weak Hamiltonian
(see \cite{Ali:1995bi} and references therein) 
\beq
\mathcal{H}_{eff} \, = \, - \, \frac{4G_F}{\sqrt{2}} \lambda_t 
\sum_{j=1}^8 C_j(\mu) O_j(\mu).
\eeq
$C_j(\mu)$ is the Wilson (short-distance)
coefficient function of the operator $O_j(\mu)$ and
$\mu = \mathcal{O}(m_b)$ is a factorization
scale.
The operator basis read:
\bea
O_1 &=& \bar{c}_{Li} \gamma^\mu b_{Lj}  
\bar{s}_{Lj} \gamma_\mu c_{Li} ;
\nonumber\\
O_2 &=& \bar{c}_{Li} \gamma^\mu b_{Li}  
\bar{s}_{Lj} \gamma_\mu c_{Lj} ;
\nonumber\\
O_3 &=& \bar{s}_{Li} \gamma^\mu b_{Li}  
\left[
\bar{u}_{Lj} \gamma_\mu u_{Lj}
+ \cdots +
\bar{b}_{Lj} \gamma_\mu b_{Lj}
\right];
\nonumber\\
O_4 &=& \bar{s}_{Li} \gamma^\mu b_{Lj}  
\left[
\bar{u}_{Lj} \gamma_\mu u_{Li}
+ \cdots +
\bar{b}_{Lj} \gamma_\mu b_{Li}
\right];
\nonumber\\
O_5 &=& \bar{s}_{Li} \gamma^\mu b_{Li}  
\left[
\bar{u}_{Rj} \gamma_\mu u_{Rj}
+ \cdots +
\bar{b}_{Rj} \gamma_\mu b_{Rj}
\right];
\nonumber\\
O_6 &=& \bar{s}_{Li} \gamma^\mu b_{Lj}  
\left[
\bar{u}_{Rj} \gamma_\mu u_{Ri}
+ \cdots +
\bar{b}_{Rj} \gamma_\mu b_{Ri}
\right];
\nonumber\\
O_7 &\equiv& \frac{e}{16 \, \pi^2} \, 
\bar{s}_i \,
\sigma^{\mu\nu} \left( m_b \, R \, + \, m_s \, L \right)
b_i \, F_{\mu\nu};
\nonumber\\
O_8 &\equiv& \frac{g_s}{16 \, \pi^2} \, 
\bar{s}_i (t^a)_{ij}
\sigma^{\mu\nu} \left( m_b \, R \, + \, m_s \, L \right)
b_j \, G_{\mu\nu}^a;
\eea
where $i,j=1,2,3$ are color indices
(fundamental $SU(3)_c$ representation).
The operators $O_1,\cdots,O_6$ are 6-dimensional
4-fermion operators, which induce $b \to s$ transitions
after contracting the repeated quark fields,
such as for example $\bar{c}_{Li}$ and $c_{Li}$
in $O_1$.
These contractions therefore generate $b \to s$ effective non-local operators,
with photons and/or gluons in the final states attached to the quark loop.
Finally $O_7/O_8$ is a 5-dimensional local 
magnetic/chromo-magnetic operator.

In this section we generalize the evaluation of the
photon spectrum
by including all the operators in $\mathcal{H}_{eff}$.
Let us first omit from our computations the operator $O_8$, which has rather
peculiar properties to be discussed later.


\subsubsection{Massless case}

Let us first summarize the results in the
simpler massless case (\cite{Aglietti:2001br} and
references therein), in a notation which easily generalizes
to the massive case. 

Let's first consider the rare $b$ decay into the simplest
final state, namely:
\beq
b \, \to \, s \, + \, \gamma.
\eeq
At lowest order, $\mathcal{O}(\alpha_{em})$, 
only the magnetic operator $O_7$
and some 4-fermion operators give a non-zero
contribution. 
An important point is that the effect of these 
4-fermion operators can be absorbed into
a redefinition of the Wilson coefficient function of $O_7$ 
(\cite{Ali:1995bi} and references therein):
\beq
C_7 \quad\mapsto\quad  
C_7^{\mathrm{eff}} \, \equiv \, 
C_7 \, + \, e_d \, C_5 \, + \, 3 \, e_d \, C_6;
\eeq
where $e_d=-1/3$ is the electric charge of a down-type quark
(in units of the proton charge).
The lowest-order rate reads:
\beq
\Gamma^{(0)}_{77} \, = \, \frac{G_F^2 m_b^5}{32 \, \pi^3} \, 
\big(C_7^{\mathrm{eff}}\big)^2 \, \left| \lambda_t \right|^2
\, \frac{\alpha_{em}}{\pi}. 
\eeq
Let's now consider the rare $b$ decay with an additional real gluon in the final state:
\beq
b \, \to \, s \, + \, g \, + \, \gamma.
\eeq
The second important point is that {\it only}
the contribution to the rate from 
$O_7$ 
(schematically, the term $\langle O_7 \rangle \langle O_7 \rangle$),
contains non-integrable infrared singularities.
All the other contributions to the partial rate,
let's call them
\beq
\Gamma_{ij}^{(1)}(y) \equiv 
\int\limits_0^y \frac{d\Gamma_{ij}^{(1)}}{dy'} \, dy', 
\eeq
such as for example $\Gamma_{27}^{(1)}(y)$
or $\Gamma_{22}^{(1)}(y)$,
contain integrable infrared singularities
or are finite.
In other words, $\Gamma_{ij}^{(1)}(y)$ for
$(i,j) \ne (7,7)$
does not contain any $\log^2 y$ and $\log y$ terms without power-suppressed coefficients.
Therefore these terms only contribute to the 
$\mathcal{O}(\alpha_S)$
Coefficient and Remainder functions
in the resummed formulae.
The partially integrated rate is then of the form:
\bea
\int\limits_0^y \frac{d\Gamma}{dy'}(y',a) \, dy'
&=& \Gamma^{(0)}_{77}
\left\{
1 \, + \, a \, K_{77}^{(1)}  
\, - \, \frac{a}{2} \, \log^2(y) 
\, - \, a \, \frac{7}{4} \log(y)
\, + \, a \, \mathrm{Rem}_{77}^{(1)}(y) +
\right.
\nonumber\\
&& \left. \qquad\qquad 
 + \, \sum_{(i,j) \ne (7,7),(8,8)} 
 a \, F_{ij}^{(1)}(y)
\right\} 
\, + \, \mathcal{O}(a^2); 
\eea
where $K_{77}^{(1)}$ is a constant whose
actual value, as we are going to
show in a moment, is immaterial, 
the $O_7O_7$-Remainder function $\mathrm{Rem}_{77}^{(1)}(y)$
is the function $H_0^{(1)}(y)$ in eq.(\ref{eq_H1}),
and
\beq
F_{ij}^{(1)}(y) \, \equiv \,  
\frac{\Gamma_{ij}^{(1)}(y)}{\Gamma^{(0)}_{77}},
\qquad (i,j) \, \ne \, (7,7), \, (8,8).
\eeq
By defining:
\beq
K_{ij}^{(1)} \, \equiv \, F_{ij}^{(1)}(y=0),
\qquad (i,j) \, \ne \, (7,7), \, (8,8),
\eeq
and
\beq
\mathrm{Rem}_{ij}^{(1)}(y)  \, \equiv \, 
F_{ij}^{(1)}(y) \, - \, F_{ij}^{(1)}(0),
\qquad (i,j) \, \ne \, (7,7), \, (8,8),
\eeq
the partial rate above is rewritten:
\bea
\int\limits_0^y \frac{d\Gamma}{dy'}(y';a) \, dy'
\, = \, 1 &+& a \, \sum_{(i,j) \ne (8,8)} K_{ij}^{(1)}  
\, - \, \frac{a}{2} \, \log^2(y) 
\, - \, a \, \frac{7}{4} \log(y) \, +
\nonumber\\
&+&  a \, \sum_{(i,j) \ne (8,8)} \mathrm{Rem}_{ij}^{(1)}(y).
\eea
By dividing by the total rate,
\beq
\int\limits_0^1 \frac{d\Gamma}{dy}(y;a) \, dy
\, = \, 1 \, + \, a \, \sum_{(i,j) \ne (8,8)} K_{ij}^{(1)}  
\, + \, a \, \sum_{(i,j) \ne (8,8)} \mathrm{Rem}_{ij}^{(1)}(y=1),
\eeq
we obtain the event fraction $E_0 \equiv E_{m_s=0}$ to $\mathcal{O}(\alpha_S)$, which is naturally factorized as (compare with eq.(\ref{eq_resum_first_ord})):
\beq
\label{EF_meq0}
E_0(y;a) \, = \, C_0(a) \, \Sigma_0(y,a) \, + \, \mathrm{Rem}_0(y,a),
\eeq
where:
\beq
C_0(a) \, \equiv \, 1 \, - \, a \sum_{(i,j) \ne (8,8)} 
\mathrm{Rem}_{ij}^{(1)}(y=1)
\, + \, \mathcal{O}(a^2),
\eeq
$\Sigma_0(y;a)$ is given in eq.(\ref{QCD_FF_massless_ord1}) and
\beq
\mathrm{Rem}_0(y,a) \, \equiv \, 
a \sum_{(i,j) \ne (8,8)} \mathrm{Rem}_{ij}^{(1)}(y) 
\, + \, \mathcal{O}(a^2).
\eeq
The following remarks are in order:
\begin{enumerate}
\item
All the constants $K_{ij}^{(1)}$
cancel by dividing the partial rate by the total
rate;
\item
As far as the dependence on the Wilson coefficient functions is concerned:
\beq
\mathrm{Rem}_{ij}^{(1)}(y) \, \propto \, \frac{C_i \, C_j}{
\big( C_7^{\mathrm{eff}} \big)^2 }.
\eeq
\end{enumerate}


\subsubsection{Massive case}

Let us now consider the factorization of the photon spectrum
in the massive case, $m_s \ne 0$, which is our primary concern.
In the massive case, the lowest-order rate 
gets a correction factor $(1+r)(1-r)^3=(1+2\rho)/(1+\rho)^4$,
so that:
\beq
\Gamma^{(0)}_{77}(\rho ) \, = \, \frac{G_F^2 m_b^5}{32 \, \pi^3} \, 
\big(C_7^{\mathrm{eff}}\big)^2 \, \left| \lambda_t \right|^2
\, \frac{\alpha_{em}}{\pi} 
\frac{1+2\rho}{(1+\rho)^4}. 
\eeq
Since, as far as we know, no analytic expressions
of the $\Gamma_{ij}^{(1)}(y,\rho)$
contributions of the photon spectrum are available
at present in the massive case, except for $i=j=7$
\footnote{
In ref.\cite{Ali:1990tj} a one-dimensional integral representation
(over the gluon energies)
of the contributions to the differential photon spectrum
coming from $\langle O_2 \rangle \langle O_7 \rangle$
and $\langle O_2 \rangle \langle O_2 \rangle$
is given.
In this paper it is also shown that the $O_1$ contribution
vanishes.
}, 
let us present a general discussion along the lines
of the massless factorization. 
Similarly to the massless case,
the partially integrated rate can be written
in the form:
\bea
\int\limits_0^y \frac{d\Gamma}{dy'}(y',\rho,a) \, dy'
&=& \Gamma^{(0)}_{77}(\rho)
\left\{
1 \, + \, a \, K_{77}^{(1)}(\rho)
\, + \, a \, \Sigma^{(1)}(y,\rho)
  \, + \, a \, \mathrm{Rem}_{77}^{(1)}(y;\rho) \, +
\right.
\nonumber\\
&& \left. \qquad\qquad 
+ \, \sum_{(i,j) \ne (7,7),(8,8)} a \, F_{ij}^{(1)}(y;\rho)
\right\}
\, + \, \mathcal{O}(a^2); 
\eea
where $K_{77}^{(1)}(\rho)$ is a  
$\rho$-dependent constant,  
$\Sigma^{(1)}(y,\rho)$ is given in eq.(\ref{eq_gen_FF_expand_1}),
$\mathrm{Rem}_{77}^{(1)}(y;\rho)$
is given in eq.(\ref{eq_Rem1_massive}) and
\beq
F_{ij}^{(1)}(y;\rho) \, \equiv \,  
\frac{\Gamma_{ij}^{(1)}(y,\rho)}{\Gamma^{(0)}_{77}(\rho)},
\qquad (i,j) \, \ne \, (7,7), \, (8,8).
\eeq
These quantities have the corresponding massless limits above:
\beq
\lim_{\rho \to 0^+} K_{77}^{(1)}(\rho) \, = \, K_{77}^{(1)};
\qquad
\lim_{\rho \to 0^+} F_{ij}^{(1)}(y;\rho) \, = \,
F_{ij}^{(1)}(y).
\eeq
By defining:
\beq
K_{ij}^{(1)}(\rho) \, \equiv \, F_{ij}^{(1)}(y=0;\rho)
\eeq
and
\beq
\mathrm{Rem}_{ij}^{(1)}(y;\rho)  \, \equiv \, 
F_{ij}^{(1)}(y;\rho) \, - \, F_{ij}^{(1)}(y=0;\rho)
\qquad (i,j) \, \ne \, (7,7), \, (8,8),
\eeq
the partial rate is written:
\bea
\int\limits_0^y \frac{d\Gamma}{dy'}(y',\rho,a) \, dy'
\, = \, 1 &+& a \, \sum_{(i,j) \ne (8,8)} K_{ij}^{(1)}(\rho)  
\, + \, a \, \Sigma^{(1)}(y,\rho) \, +
\nonumber\\
&+&  a \, \sum_{(i,j) \ne (8,8)} \mathrm{Rem}_{ij}^{(1)}(y;\rho)
\, + \, \mathcal{O}(a^2).
\eea


\subsubsection{General Factorization Scheme}

By dividing the partial rate by the total rate,
we obtain the event fraction $E$ to $\mathcal{O}(\alpha_S)$, which has a general factorization of the form:
\beq
\label{EF_mneq0}
E(y;\rho,a) \, = \, C(\rho,a) \, \Sigma(y,\rho,a) \, + \, \mathrm{Rem}(y,\rho,a),
\eeq
where:
\beq
\label{eq_Coef_Fun_Gen_gen}
C(\rho,a) \, \equiv \, 1 \, - \, a \sum_{(i,j) \ne (8,8)} 
\mathrm{Rem}_{ij}^{(1)}(y=1;\rho)
\, + \, \mathcal{O}(a^2),
\eeq
$\Sigma(y,\rho,a)$ has been given to $\mathcal{O}(a)$ 
in eq.(\ref{eq_gen_FF_expand_1})
and
\beq
\label{eq_Rem_Fun_Gen_gen}
\mathrm{Rem}(y,\rho,a) \, \equiv \, 
a \sum_{(i,j) \ne (8,8)} \mathrm{Rem}_{ij}^{(1)}(y;\rho) 
\, + \, \mathcal{O}(a^2).
\eeq


\subsubsection{Improved Factorization Scheme}

Let us now describe the Improved Factorization Scheme
in the general case.
The terms in the massive Remainder function are naturally decomposed into a commuting piece
and a non-commuting one, namely:
\beq
\label{eq_splitting}
\mathrm{Rem}_{ij}^{(1)}(y;\rho) 
\, = \, L_{ij}^{(1)}(y;\rho) \, + \, 
N_{ij}^{(1)}(y;\rho),
\eeq
where:
\beq
\lim_{\rho \to 0^+} 
\left[
\lim_{y \to 0^+} L_{ij}^{(1)}(y;\rho)
\right]
\, = \, 
\lim_{y \to 0^+} 
\left[
\lim_{\rho \to 0^+} L_{ij}^{(1)}(y;\rho)
\right],
\eeq
while:
\beq
\label{eq_split_termn_NC}
\lim_{\rho \to 0^+} 
\left[
\lim_{y \to 0^+} N_{ij}^{(1)}(y;\rho)
\right]
\, \ne \, 
\lim_{y \to 0^+} 
\left[
\lim_{\rho \to 0^+} N_{ij}^{(1)}(y;\rho)
\right].
\eeq
The splitting in eq.(\ref{eq_splitting}) can be made on a term-by-term
analysis of the Remainder function contributions, either analytically or numerically.
The commuting terms do not pose any problem and
are treated as in the general factorization scheme above.
If $L_{ij}(y=0;\rho) \ne 0$, this term gives a contribution 
$L_{ij}(y=0;\rho)$ to the Improved Coefficient function and a contribution
$L_{ij}(y;\rho)-L_{ij}(y=0;\rho)$ to the
Improved Remainder function.

Let us now consider the (more complicated) 
non-commuting terms. We can assume that:
\beq
\label{eq_vanishing_Nij}
\lim_{y \to 0^+} N_{ij}^{(1)}(y;\rho) \, = \, 0. 
\eeq
If that is not the case, we impose the above limit by
simply subtracting from $N_{ij}^{(1)}(y;\rho)$ 
its value at $y=0$:
\beq
N_{ij}^{(1)}(y;\rho) \quad \mapsto \quad 
N_{ij}^{(1)}(y;\rho) \, - \, N_{ij}^{(1)}(0;\rho).
\eeq
The constant $N_{ij}^{(1)}(0;\rho)$ is then added back to the
commuting contributions (being independent on $y$,
it is trivially a commuting term).

Now the idea is simply to treat the non-commuting terms just like the term $y/(y+\rho)$ above, by means of the Partition of Unity,
so that:
\begin{enumerate}
\item
The $\mathcal{O}(\alpha_S)$ Improved Coefficient function is obtained by
adding to the General Coefficient function 
$C^{(1)}(\rho)$ above, eq.(\ref{eq_Coef_Fun_Gen_gen}),
the terms
$\Phi_\Delta(y-\rho) \, N_{ij}^{(1)}(y;\rho)$,
giving:
\beq
C_I^{(1)}(y;\rho) \, = \,
 - \, \sum_{(i,j) \ne (8,8)} 
\mathrm{Rem}_{ij}^{(1)}(y=1;\rho)
\, + \, \Phi_\Delta(y-\rho )\sum_{(i,j) \ne (8,8)} 
N_{ij}^{(1)}(y;\rho);
\eeq
\item
The $\mathcal{O}(\alpha_S)$ Improved Remainder function 
is obtained from  the General one, 
$\mathrm{Rem}^{(1)}(y;\rho)$ in eq.(\ref{eq_Rem_Fun_Gen_gen}),
by replacing the non-commuting terms
$N_{ij}^{(1)}(y;\rho)$ with the terms 
$\Phi_\Delta(\rho-y) \, N_{ij}^{(1)}(y;\rho)$
respectively, giving:
\beq
\mathrm{Rem}_I^{(1)}(y;\rho)
\, = \,
\sum_{(i,j) \ne (8,8)} 
\left[
L_{ij}^{(1)}(y;\rho)
\, + \, \Phi_\Delta(\rho-y) \, N_{ij}^{(1)}(y;\rho)
\right].
\eeq 
\end{enumerate}
%
By using eq.(\ref{eq_fundam_relat}), 
it is immediate to check that
\beq
\Phi_\Delta(y-\rho) \, N_{ij}^{(1)}(y;\rho) \, + \,
\Phi_\Delta(\rho-y) \, N_{ij}^{(1)}(y;\rho) \, = \,
N_{ij}^{(1)}(y;\rho),
\eeq
so that the $\mathcal{O}(\alpha_S)$ event fraction
is correctly reproduced.


\subsubsection{Alternative formulation}

As in the previous explicit computation, 
the $y$-dependent contribution
to the Improved Coefficient function can be simplified 
by taking the massless limit $\rho \to 0^+$ in
$N_{ij}^{(1)}(y;\rho)$, giving:
\beq
\label{CF_gen_simpl}
C_I^{(1)}(y;\rho) \, = \,
 - \, \sum_{(i,j) \ne (8,8)} 
\mathrm{Rem}_{ij}^{(1)}(y=1;\rho)
\, + \, \Phi_\Delta(y-\rho )\sum_{(i,j) \ne (8,8)} 
N_{ij}^{(1)}(y;\rho=0).
\eeq
That generalizes the partial
fractioning made in the $O_7\,O_7$-case
\beq
\frac{y}{y + \rho} \, = \, 1 \, - \, \frac{\rho}{y + \rho}.
\eeq
The Improved Remainder function corresponding to the 
Improved Coefficient Function given in eq.(\ref{CF_gen_simpl}) then reads:
\beq
\mathrm{Rem}_I^{(1)}(y;\rho)
\, = \,
\sum_{(i,j) \ne (8,8)} 
\left[
L_{ij}^{(1)}(y;\rho)
\, +  \, N_{ij}^{(1)}(y;\rho)
\, - \, 
\Phi_\Delta(y-\rho) \, N_{ij}^{(1)}(y;\rho=0)
\right].
\eeq
It is easy to check that also the function
on the r.h.s. above
vanishes for $y \to 0^+$, as it should.
Just remember eq.(\ref{eq_vanishing_Nij}) 
and the definition of the function
$\Phi_\Delta(x)$, eq.(\ref{eq_def_Phi_Delta}), 
together with the fact that $\Delta < \rho$
(note that $N_{ij}^{(1)}(y;\rho=0)$ does not
vanish for $y \to 0^+$ because of eq.(\ref{eq_split_termn_NC})).


\subsubsection{Double insertion of $O_8$}

Let us now consider the effects of the operator 
$O_8$ in the radiative decay $B \to X_s \gamma$. 
This operator is obtained from $O_7$
simply by replacing the electromagnetic field strength
$F_{\mu\nu}$ by the $QCD$ one, $G_{\mu\nu}^a t^a$,
as well as by replacing the electric charge $e$ by the color charge
$g_s$ ($\alpha_{em} \equiv e^2/(4\pi), \,\, \alpha_S \equiv g_s^2/(4\pi)$).
As a consequence, the lowest-order matrix element
of $O_8$ induces the decay
\beq
\label{eq_tree_O8}
b \, \to \, s \, + \, g.
\eeq
Note that the gluon and the strange quark are emitted 
locally by $O_8$.
There are no photons in the final state, 
which experimentally consists,
in the beauty rest frame, of {\it two} back-to-back hadronic jets.
The topology of the final states in the decay (\ref{eq_tree_O8}) is quite different from that of the tree-level
decay mediated by $O_7$,
\beq
b \, \to \, s \, + \, \gamma,
\eeq
which consists of {\it one} hadronic jet recoiling against a
(high-energy) photon.

The lowest-order contribution of $O_8$ to the 
differential photon spectrum
is then a spike at vanishing photon energy:
\beq
\frac{d \Gamma^{(0)}_{88}}{dx}(x,\rho)
\, = \, \Gamma^{(0)}_{88}(\rho)  \, \delta(x),
\eeq
where:
\beq
\Gamma^{(0)}_{88}(\rho) \, = \, \frac{G_F^2 m_b^5}{32 \pi^3} 
\left| C_8^{\mathrm{eff}} \lambda_t \right|^2
\frac{C_F \alpha_S}{\pi} \, \frac{1+2\rho}{(1+\rho)^4}. 
\eeq
$C_8^{\mathrm{eff}}$ is an effective (or improved) Wilson coefficient function of $O_8$, including 
the (non-vanishing) contributions from the 4-fermion
operators \cite{Ali:1995bi}: 
\beq
C_8^{\mathrm{eff}} \, = \, C_8 \, + \, C_5.
\eeq
Note that $\Gamma^{(0)}_{88}(\rho)$ is obtained from
$\Gamma^{(0)}_{77}(\rho)$ above simply replacing
$C_7^{\mathrm{eff}}$ with $C_8^{\mathrm{eff}}$
and $\alpha_{em}$ with $C_F \alpha_S$.

Let us now consider the $\mathcal{O}(\alpha_{em})$
corrections to the decay (\ref{eq_tree_O8}) involving a real photon, 
contributing to the process
\beq
b \, \to \, s \, + \, g \, + \, \gamma.
\eeq
The gluon is again emitted locally by $O_8 $, while the photon is emitted by the beauty and the strange
quark lines.
The contribution to the differential photon spectrum reads \cite{Ali:1995bi}:
\beq
\frac{1}{\Gamma^{(0)}_{88}} \frac{d \Gamma^{(1R)}_{88}}{dx}
\, = \, \frac{\alpha_{em} }{\pi} \, e_d^2 \, X^{(1R)},
\eeq
where $e_d=-1/3$ and:
\bea
X^{(1R)} &=&
\frac{1}{2 \, (1-r)}
\left[
\frac{1 + (1-x)^2}{x} 
\, + \, r \, \frac{2 + x(2-x)}{x}
\right]
\left[
\log \frac{1}{r} \, + \, \log(1-x + r \, x)
\right] \, + \qquad
\\
&-& \frac{2 \, (1-x)}{x}
\, + \, \frac{1-r}{4} \left( \frac{1-x}{1-x+rx} \right)^2 (1-2x)
\, - \, \frac{1-r}{4} \frac{1-x}{1-x+rx} 
\left( 1-x+2x^2 \right).
\nonumber
\eea
The above spectrum contains a mass singularity of collinear
origin for $m_s\to 0^+$, 
as well as a soft singularity for vanishing photon energy
($x \to 0^+$):
\beq
\label{eq_spec_O8O8}
\frac{1}{ \Gamma^{(0)}_{88} } \frac{ d \Gamma_{88}^{(R)} }{dx}
\, \simeq \, 
\delta(x)
\, + \, e_d^2 \, \frac{\alpha_{em} }{2 \pi} \, 
\hat{P}^{(0)}_{\gamma e}(x) \log\left( \frac{m_b^2}{m_s^2} \right)
\, - \, e_d^2 \,\frac{\alpha_{em} }{\pi} \,  \frac{2}{x},
\qquad 0 < x, r \ll 1,
\eeq
where $\hat{P}^{(0)}_{\gamma e}(x)$
is the leading-order unregularized $QED$ splitting function
of an electron (or a positron) into a photon:
\beq
\hat{P}^{(0)}_{\gamma e}(x) \, = \, \hat{P}^{(0)}_{ee}(1-x) 
\, = \, \frac{1 + (1-x)^2}{x}.
\eeq
By coupling the quarks to the electromagnetic field,
the strange quark produces a $QED$ jet,
as the leading contribution on the r.h.s. of eq.(\ref{eq_spec_O8O8}) consists of a soft photon emitted at a small angle
with respect to the strange quark motion direction.
Note that the topology of the $b \to s + g + \gamma$ final states
mediated by $O_8$ involves two back-to-back
jets, initiated by the strange quark and by the gluon.
The jet initiated by the strange quark also contains
the detected photon. Experimentally, a large hadronic activity around the final photon
is then expected.
The topology of the $b \to s + g + \gamma$ final states
mediated by $O_7$ is quite different,
as it involves {\it one} hadronic jet containing, 
to $\mathcal{O}(\alpha_S)$,
the strange quark and the gluon, recoiling against
the (hard) photon. 
In the latter, $O_7O_7$-case, the photon is then
expected to be isolated.

The r.h.s. of eq.(\ref{eq_spec_O8O8}) also contains a single-logarithmic term $\propto 1/x$ (upon integration over $x$),
coming from soft, not collinearly enhanced,
radiation off the beauty and the strange
quarks (the factor {\it two} comes indeed from having
{\it two} massive quarks in the process). 
This soft radiation is roughly isotropic in space (in beauty rest frame)
and is then not naturally associated to any jet;
it represents a kinematic violation of 
independent jet fragmentation --- the latter coming from angular ordering%
\footnote{
Angular ordering is, in turn, an approximate
consequence of color coherence --- a fundamental
dynamical property of perturbative QCD. 
} ---
at the next-to-leading level%
\footnote{
In process involving light quarks or gluons only,
such as for example Drell-Yan production of
intermediate vector bosons or Higgs production
by gluon-gluon fusion, independent jet fragmentation
is instead violated {\it dynamically}, usually
{\it one order further},
i.e. at next-to-next logarithmic accuracy.
}.
As well known, the main effect of the $\mathcal{O}(\alpha_{em})$ virtual corrections to the $O_8O_8$ tree-level decay,
is to introduce a plus regularization
in the splitting function and in the soft-singular function,
so that:
\beq
\frac{1}{ \Gamma^{(0)}_{88} } \, \frac{ d \Gamma^{(R+V)} }{dx}
\, \simeq \, 
\delta(x)
\, + \, e_d^2 \, \frac{\alpha_{em} }{2 \pi} \,  
P^{(0)}_{\gamma e}(x) \log\left( \frac{m_b^2}{m_s^2} \right)
\, - \, 2 \, e_d^2 \, \frac{\alpha_{em} }{\pi} \,  
\left( \frac{1}{x} \right)_+ ,
\eeq
where:
\beq
P^{(0)}_{\gamma e}(x) \, = \, 
\left[
\frac{1+(1-x)^2}{x} 
\right]_+
\equiv 
\frac{1+(1-x)^2}{x} 
\, - \, \delta(x) \int\limits_0^1 \frac{1+(1-y)^2}{y} \, dy 
\eeq
and
\beq
\left( \frac{1}{x} \right)_+ \, \equiv \,
\frac{1}{x} \, - \, \delta(x) \int\limits_0^1 \frac{dy}{y}.
\eeq
By adding virtual photon corrections,
soft singularities cancel 
(in a distribution sense in the differential distribution),
while collinear singularities, for $m_s \to 0^+$, do not. 
Therefore one has to factorize the $QED$ collinear logarithm
above
by means of an ad-hoc fragmentation function,
$D_{\gamma s}(x,Q^2)$. 
The latter is a universal, i.e. process-independent,
function, which can be interpreted as the probability of finding
a photon inside a jet initiated by a strange quark,
with a fraction $x$ of the initial strange energy,
in a process with hard scale $Q$ ($Q=m_b$ in our case). 
Since the strange quark mass is of the order of the $QCD$
scale, 
\beq
m_s \, \approx \, \Lambda_{QCD},
\eeq
substantial non-perturbative corrections are expected.
In order to avoid the introduction of
a non-perturbative function --- leading in real life to a loss
of predictivity --- ,
one can modify the definition of the observed final states, by requiring,
for example, the photon to be angularly isolated, in some way, from the final
partons/hadrons in the event.  

Finally, let us remark that the soft-photon region,
$x \ll 1$, where the operator $O_8$ dominates,
is experimentally not interesting
due to the huge background.


\section{Conclusions}
\label{sec_conclus}

We have considered various factorization
schemes for threshold resummation
in processes involving the decay
of a heavy quark into a different heavy (massive) quark,
accompanied by non-colored partons,
i.e. in practice photons, leptons or vector bosons.

By taking the radiative $B \to X_s \gamma$ decay as a model process
and restricting ourselves to the leading operator $O_7$
in the effective $b \to s \gamma$ weak Hamiltonian,
we have first considered soft gluons only
and we have constructed  a simple
soft factorization scheme. 
The latter can be consistently applied to the heavy quark decays 
so long as the ordinary velocity of the
final quark,
in the initial quark rest frame,
is not too large 
(compared to light velocity $c$),
so that collinear effects 
(collinear logarithms) are not large.
The soft scheme can be probably applied to
the CKM-favored semileptonic $B$ decays, 
\beq
\label{eq_dec_semilep_CKMfavor}
B \, \to \, X_c \, + \, l \, + \, \nu_l,
\eeq
as the charm ordinary 3-velocity $v_c$ never becomes too large
in this case:
\beq
v_c \, \lsim \, 0.77 \, c.
\eeq
We have then constructed a general
massive factorization scheme,
which correctly works for a non-zero
(and not too small) final quark mass $m_s$.
However, we have found that this scheme
does not behave well in the massless limit 
$m_s \to 0^+$,
because its Coefficient function and
its Remainder function do not approach, in this limit,
the corresponding functions of the 
standard factorization formula constructed {\it after} 
taking the massless limit of the photon spectrum.
We have shown that this mismatch is generated by a simple 
term in the photon spectrum (equivalent to the distribution in 
the final hadron invariant mass squared $m_{X_s}^2$),
namely the term, in units of $C_F \alpha_S/(4\pi)$,
\beq
\frac{m_s^2 \, - \, m_{X_s}^2}{ m_{X_s}^2 }.
\eeq
Indeed, for the above term, 
the massless limit $m_s \to 0^+$ 
and the threshold limit $m_{X_s} \to m_s^+$
do not commute with each other.
In terms of the mass-correction parameter 
$\rho \equiv m_s^2/(m_b^2-m_s^2)$
and the threshold variable 
$y \equiv (m_{X_s}^2-m_s^2)/(m_b^2-m_s^2)$,
the above term has been written in the 
main body of the paper as
\beq
- \, \frac{y}{y \, + \, \rho}.
\eeq
In the new variables,
it is immediate to check that
the massless limit $\rho \to 0^+$ and 
the threshold limit $y \to 0^+$
do not commute with each other.
It is natural to expect the appearance of 
{\it such} terms in the photon spectrum (for which the 
limits $y \to 0^+$ and $\rho \to 0^+$
do not commute with each other) 
to be {\it generic} and not restricted
to the $O_7$ operator.
A general discussion of the effects, 
in the $B \to X_s \gamma$ photon spectrum,
of all the subleading operators in the effective weak 
Hamiltonian has also been presented.

Since in semileptonic $b \to c$ decays, 
eq.(\ref{eq_dec_semilep_CKMfavor}),
the heavy quark mass ratio is rather 
large, $m_c/m_b \approx 1/3$,
we expect the general massive scheme
to be consistently applied to describe them.
We also expect the massive scheme and the soft
scheme to give quantitatively similar results
for these decays.


Finally, we have constructed an Improved
factorization scheme for the massive case, $m_s \ne 0$,
which has the correct massless limit $m_s \to 0^+$.
That is the main result of our work.
A main point is that, to that aim, 
we have been forced to introduce 
in the Improved Coefficient function $C_I$, 
in addition to the standard dependence on 
the mass-correction parameter 
$\rho$, also a dependence on the threshold 
variable $y$:
\beq
C_I \, = \, C_I(\rho,y).
\eeq 
Actually, we constructed an Improved Coefficient function
which has a smooth dependence both on $y$ and $\rho$,
and which is close to the massless Coefficient function
in the quasi-massless region $y \gg \rho$.
The mathematical tool we needed is the so-called 
Partition of Unity.

By means of the Improved Factorization scheme, 
it is  possible to describe, with a unique formalism
and in a smooth way, both the quasi-massless kinematic region 
$\rho \ll y$
and the pure soft region $y \ll \rho$, which are
dynamically quite different, as well
as the transition region $y \approx \rho$.
As is often the case, the interpolation 
between the asymptotic regions above,
to the slice $y \approx \rho$
is, to some extent, arbitrary, as many different
functions can be used to that aim.
In other words, there is an ambiguity,
in the interpolation from the region
$\rho \ll y$ to the region $y \ll \rho$, 
which is never solved in perturbation
theory, but only shifted formally to higher orders.

The improved factorization scheme can
be applied, for example, to $CKM$-favored top 
quark decays $t \to X_b + W$, 
in the study of
the final hadron invariant squared mass distribution
in the intermediate or transition region
\beq
m_{X_b}^2 \, - \, m_b^2 \, \approx \, m_b^2 \, \ll \, m_t^2,
\eeq
as well as in the soft region
\beq
m_{X_b}^2 \, - \, m_b^2 \, \ll \, m_b^2
\, \ll \, m_t^2
\eeq
and in the effectively-massless region
\beq
m_b^2 \, \ll \, m_{X_b}^2
\, \ll \, m_t^2.
\eeq
The Improved scheme
can also be applied to more complicated
decays, such as for example the semileptonic
$b \to c$ decay, eq.(\ref{eq_dec_semilep_CKMfavor}) 
--- a three-body decay at tree level.

We believe that our scheme
can be generalized to all the hard processes in which one
observes the hadron invariant mass $m_X$
of a jet $X$ initiated by a heavy quark $Q$
in all the possible kinematic regions,
namely $m_X - m_Q \ll m_Q$, $m_X - m_Q \approx m_Q$ and $m_X \gg m_Q$.


We have explicitly constructed the Improved factorization scheme 
at order $\alpha_S$, i.e. at Next-to-Leading Logarithmic
($NLL$) accuracy.
It would be interesting to explicitly extend
the scheme to the next order, i.e. with
$\mathcal{O}(\alpha_S^2)$ Coefficient
and Remainder functions.
The idea of using the Partition of Unity
could also be generalized to higher orders.

An extension of our scheme in a different direction could be
the construction of an improved factorization formula 
for resummed transverse momentum distributions 
involving heavy quarks. 


\end{document}